\documentclass[12pt,preprint]{aastex}

\newcommand{\eg}{{\it e.g.}}
\newcommand{\etal}{et~al.}
\newcommand{\hk}{$H-K$}

\newcommand{\ik}{$I-K$}

\newcommand{\lbol}{$L_{\rm bol}$}

\newcommand{\lx}{$L_{\rm x}$}
\newcommand{\mdot}{$\dot{M}$}
\newcommand{\msun}{M$_{\sun}$}

\newcommand{\teff}{$T_{\rm eff}$}

\newcommand{\uv}{$U-V$}

\newcommand{\vsini}{$v \sin i$}

\begin{document}

\title{Chandra X-Ray Observations of Young Clusters. III. NGC
2264 and the Orion Flanking Fields}

\author{L.\ M.\ Rebull, J.\ R.\ Stauffer}
\affil{Spitzer Science Center/Caltech, M/S 220-6, 1200 E.\ California Blvd.,
Pasadena, CA  91125 (luisa.rebull@jpl.nasa.gov)}

\and
\author{S.\ V.\ Ramirez}
\affil{IPAC/Caltech, M/S 100-22, 1200 E.\ California Blvd.,
Pasadena, CA  91125}

\and

\author{E.\ Flaccomio, S.\ Sciortino, G.\ Micela}
\affil{INAF-Osservatorio Astronomico di Palermo, Piazza del Parlamento 1,
90134 Palermo, Italy}

\and

\author{S.\ E.\ Strom, S.\ C.\ Wolff}
\affil{National Optical Astronomy Observatory, 950 North Cherry
Avenue, Tucson, AZ 85726}

\begin{abstract}

Chandra observations of solar-like pre-main sequence (PMS) stars
in the Orion Flanking Fields (age $\sim$1 Myr) and NGC 2264
($\sim$3 Myr) are compared with the results of the COUP survey
of similar objects in the ONC ($\sim$0.5 Myr).  The correlations
between log \lx\ and mass found for PMS stars on convective
tracks in these clusters are consistent with the relationships
found for the ONC, indicating little change in the {\em median}
values of either log \lx\ or log \lx/\lbol during the first
$\sim$3-5 Myr of evolution down convective tracks.  The fraction
of stars with extreme values of \lx, more than 10 times higher
than the average for a given \lbol\ or with log \lx/\lbol\
greater than the canonical saturation value of $-$2.9, is
however larger by a factor of two in the younger ONC when
compared with the Orion FF and NGC 2264.

PMS stars in NGC 2264 on radiative tracks have \lx/\lbol\ values
that are systematically lower by a factor of $\sim$10 times than
those found for stars of similar mass on convective tracks.  
The dramatic decrease in flux from convective to radiative
phases of PMS evolution is likely related to major structural
changes which influence the efficiency of magnetic field
generation and thus the level of magnetic activity.

As in the ONC, we find that stars with measured periods have, on
average, higher X-ray luminosities.  However, there is a wide
range in \lx\ and \lx/\lbol\ for both periodic and non-periodic
stars of similar mass.  Among stars with measured periods, the
level of X-ray emission does not correlate with the rotation
rate.  

For this data set, we find no statistically significant
correlation between X-ray flux and (a) the presence or absence
of circumstellar accretion disks, or (b) disk accretion rates as
assessed from ultraviolet excesses.

\end{abstract}


\section{Introduction}

Low-mass stars of all stages of evolution---from protostars to
the main sequence---emit X-rays (see, e.g., Feigelson \&
Montmerle 1999).  The relationships between the levels of X-ray 
emission and stellar properties for pre-main-sequence (PMS)
stars are, however, very different from the well established
correlations between rotation, Rossby number, and X-ray emission
seen in young main-sequence stars (e.g., Stauffer \etal\ 1994,
Micela \etal\ 1996, Pizzolato \etal\ 2003).  Several factors may
contribute to these differences. Low-mass PMS stars are fully
convective during their early evolution, and differences in
stellar structure (convective in the PMS phase vs. radiative
core, convective envelope on the ZAMS) are likely to produce
different levels of magnetic-field-related activity.  Many PMS
stars are still surrounded by accretion disks, which may 
influence X-ray emission in at least three ways.  Additional
X-ray emmission above coronal levels may be produced  in
accretion shocks or wind/jet shocks (e.g., Kastner \etal\ 2002,
2005; Guedel \etal\ 2005).  Depending on the system orientation
and the size of disk (and extent of flaring), disks or neutral
winds may absorb X-ray photons, decreasing the detected flux,
particularly at short wavelengths (e.g., Kastner \etal\ 2005;
Walter \& Kuhi 1981). The rotation speed of PMS stars could, in
principle, also affect X-ray emission.  If so, then the X-ray
properties of stars surrounded by accretion disks may differ
systematically from their counterparts that lack disks, owing to
the effects of ``disk locking'' on rotation (\eg, Shu \etal\
1987, 2000; K\"onigl 1991, 1989).  Finally, there might also be
other less direct interactions between accretion, the young
stellar object, its outer convective zone, the magnetic field
structures, and the X-ray emitting plasma, as discussed in
Preibisch \etal\ (2005).

The recent COUP project (Getman \etal\ 2005) has investigated
the detailed relationships among stellar parameters and X-ray
flux among solar-like PMS stars in the ONC (Preibisch \etal\
2005), which has an age of $\sim$0.5 Myr.  We consider here the
relationships among rotation, X-rays, and disks for several
hundred stars in the Orion Flanking Fields  ($\sim$1 Myr) and
NGC 2264 ($\sim$3 Myr).  We focus our analysis on objects of
masses $\sim$0.3 to 2.5 \msun.  

The stars in these regions are, on average, slightly older than
those in the ONC and are therefore, in combination with the COUP
data, well-suited for searching for  changes during the early
evolution of PMS stars.  Importantly, for stars of $\sim$1-2
\msun, the current sample includes stars spanning the range of
ages in which substantial structural changes take place as stars
evolve from a fully convective phase to a phase in which they
develop radiative cores and convective envelopes.  Since the
origin of the X-rays in the Sun is believed to be linked to
dynamo-driven activity, we might expect that the mechanism of
magnetic field and consequent coronal X-ray generation changes
as stars make the transition from a fully convective state to
one in which a radiative core is surrounded by a thin convective
envelope.  Consequently, we might also expect to see a change in
the X-ray properties.  

The basic presentation of the data for the Orion FF and NGC 2264
appears in other papers (Ramirez \etal\ 2004a,b; Flaccomio
\etal\ 2005a,b); we discuss here the implications of these
data.  After a brief summary of the observations 
(\S\ref{sec:obs}), we consider in some detail the completeness
and selection effects of  the sample of stars to be discussed
here. Then we compare the relationships of log \lx\ and log
\lx/\lbol\ (\S\ref{sec:lxlbol}) with mass and age for stars on
convective tracks in all three clusters and find that the
relationships are entirely consistent, thereby indicating that
there is little or no evolution in X-ray properties during the
first $\sim$5 Myr of evolution down convective tracks.  We find
clear evidence for a decrease in log \lx/\lbol\ when stars make
the transition from convective to radiative tracks
(\S\ref{sec:oldmass}).  We also discuss the sample with rotation
information (\S\ref{sec:rotation}), and find that the subset of
stars with measured periods has on average significantly higher
values of log \lx/\lbol, as was found previously for the ONC
(Preibish \etal\ 2005b).  Also in accordance with previous
studies, we find no clear trends of X-ray emission with rotation
among the sample having spot-modulated periods.  Finally, in
\S\ref{sec:disks}, we examine the relation between disks, mass
accretion rate (inferred from ultraviolet excess emission) and
X-ray luminosity; we find no clear trends.

\section{Observations, Data, and Sample Selection}
\label{sec:obs}

\subsection{Observations}

In this paper, we discuss stars detected in four Chandra/ACIS fields:
two in NGC 2264 and two in the outer Orion Nebula Region, the
so-called ``Orion Flanking Fields'' (FF; see Rebull \etal\ 2000,
Rebull 2001 for more discussion of these fields).    

The Chandra data for the Orion FF were first presented in
Ramirez \etal\ (2004b).  The two fields here correspond to
portions of fields 2 and 4 from Rebull \etal\ (2000); they are
centered on 05h35m19s, $-4\arcdeg48\arcmin15\arcsec$ and
05h35m6s, $-5\arcdeg40\arcmin48\arcsec$.  Standard CIAO
procedures were used to reduce the data. With about 48 ks
exposure time per field, 417 sources were detected in total over
both fields.  The distance modulus we assumed for Orion is 8.36
(470 pc).

The Chandra data for the northern field in NGC 2264 were first
presented in Ramirez \etal\ (2004a), in which 263 sources were
detected in 48.1 ks. The ACIS-I array was centered at 6h40m48s,
$+9\arcdeg51\arcmin$.   These data were also independently
analyzed by Sung \etal\ (2004) in an effort to identify X-ray
emitting WTTS.   The ACIS data for the southern field in NGC
2264 is discussed in Flaccomio \etal\ (2005, 2006).  The ACIS-I
array was centered at 6h40m58.1s, 
+9$\arcdeg$34$\arcmin$00.40$\arcsec$, and 420 sources were
detected in 96 ks.  Standard CIAO procedures were used to reduce
the data.  The distance modulus we assumed for NGC 2264 is 9.40
(760 pc).  

Following the approach discussed in Ramirez \etal\ (2004a,b) and
Flaccomio  \etal\ (2006), we determined upper limits for all
stars with optical data in these fields.

ACIS data have also been recently published in the ONC core
region by Feigelson \etal\ (2003b) and as part of the COUP
project (Getman \etal\ 2005, Preibisch \etal\ 2005b).  Those
much deeper ($\sim20\times$) observations of an extremely rich
and slightly younger region (Ramirez \etal\ 2004b) provide a
dataset of great statistical weight with which we can compare
our results.

\subsection{Ancillary Data and Counterparts}

In order to interpret these X-ray data in the context of young
stars, we assembled extensive catalogs of multi-wavelength
photometric and spectroscopic data for each cluster, as
discussed in greater detail in Ramirez \etal\ (2004a, b).  These
catalogs are hereafter referred to as the ancillary data
catalogs (ADC).  For NGC 2264, the bulk of the data in the ADC
comes from Rebull \etal\ (2002), as well as Park \etal\ (2000)
and the Two Micron All Sky Survey (2MASS).  The ADC also
contains preliminary proper motions from B.\ Jones; see Rebull
\etal\ (2002).  We have added to the ADC recently published data
from Lamm \etal\ (2004), Sung \etal\ (2004), and Dahm \& Simon
(2005).  For the Orion Flanking Fields, the ADC was assembled
from $\sim$30 published articles ranging from the $U$ magnitudes
in Rebull \etal\ (2000) to the $JHK$ photometry from 2MASS; see
Ramirez \etal\ (2004b).  The ADC has also been updated with more
recently published photometric data from Sicilia-Aguilar \etal\
(2005), as well as with 126 additional \vsini\ measurements and 
upper limits currently under preparation for publication (Rebull
\etal\ 2006 in prep). 

For each of the clusters, we matched each X-ray detection from
Ramirez \etal\ (2004a,b) to optical and/or NIR counterparts from
the ADC based on comparison of RA and Dec of the individual
objects combined with manual inspection of the near-IR field
around each X-ray source.

Following the same approach as in, e.g., Rebull \etal\ (2000),
using the stars' positions in the dereddened $I/V-I$
color-magnitude diagram (CMD), we calculated \lbol\ following
the precepts summarized by Hillenbrand (1997).  Similarly, we
assigned masses to the stars according to the two different sets
of PMS stellar models: (1) those  of D'Antona \& Mazzitelli
(1994, 1998; DAM), with conversion from the theoretical to the
observational plane (including calculation of \lbol) using
transformations found in Hillenbrand (1997); and (2) those of
Siess \etal\ (2000; SDF), the Z=0.02 model with no overshooting,
using conversions from Kenyon \& Hartmann (1995).  The DAM
models are what we have used in previous papers (e.g., Rebull
\etal\ 2000), but the SDF models were used by COUP  (Preibisch
\etal\ 2005b); masses from both models are discussed here to
facilitate comparison with these and other papers. If stars fell
outside the grid provided by the models, no masses or ages were
assigned; such stars were excluded from this analysis.
Additional discussion regarding uncertainties in masses and ages
appears in Rebull (2001); we repeat here that the largest
uncertainties are in the models themselves, and the ages and
masses used here are perhaps best thought of as proxies for
relative placements in the observational CMD.

In Ramirez \etal\ (2004b), we have found that the X-ray selected
sample in Orion FF ($\sim$1 Myr) is on average slightly older
than that for the ONC ($\sim$0.5 Myr), but not as old as that
for NGC 2264 ($\sim$3 Myr). Repeating this analysis now
including the second NGC 2264 field, it is still the case that
the median age of the FF sample is older than the ONC, and NGC
2264 is older still.  The {\em range} of ages found within {\em
each} of these clusters is on the order of $\sim$3 Myr and
comparable in all three clusters, with the largest {\em
fraction} of older stars being found in NGC 2264; see
Figure~\ref{fig:cmd}.  The ages listed here (0.5, 1, and 3 Myr)
are obtained from comparing distributions of ages derived using
DAM models for the ``best possible sample'' (see next subsection
below) for these clusters.   If we use SDF models for the same
subset of stars, we find ages that are on average older by 
$\sim$1 Myr, but the relative age placement of the clusters is
similar at 1.5, 2, and 4 Myr.  The relative placement of the
clusters is also consistent with the relative numbers of
radiative and convective stars (see below); the fewest radiative
stars are found in the ONC and the most are found in NGC 2264.

\subsection{Catalog Statistics and Sample to be Considered Here}
\label{sec:bestsample}


\begin{figure*}[tbp]
\epsscale{1.0}
\plottwo{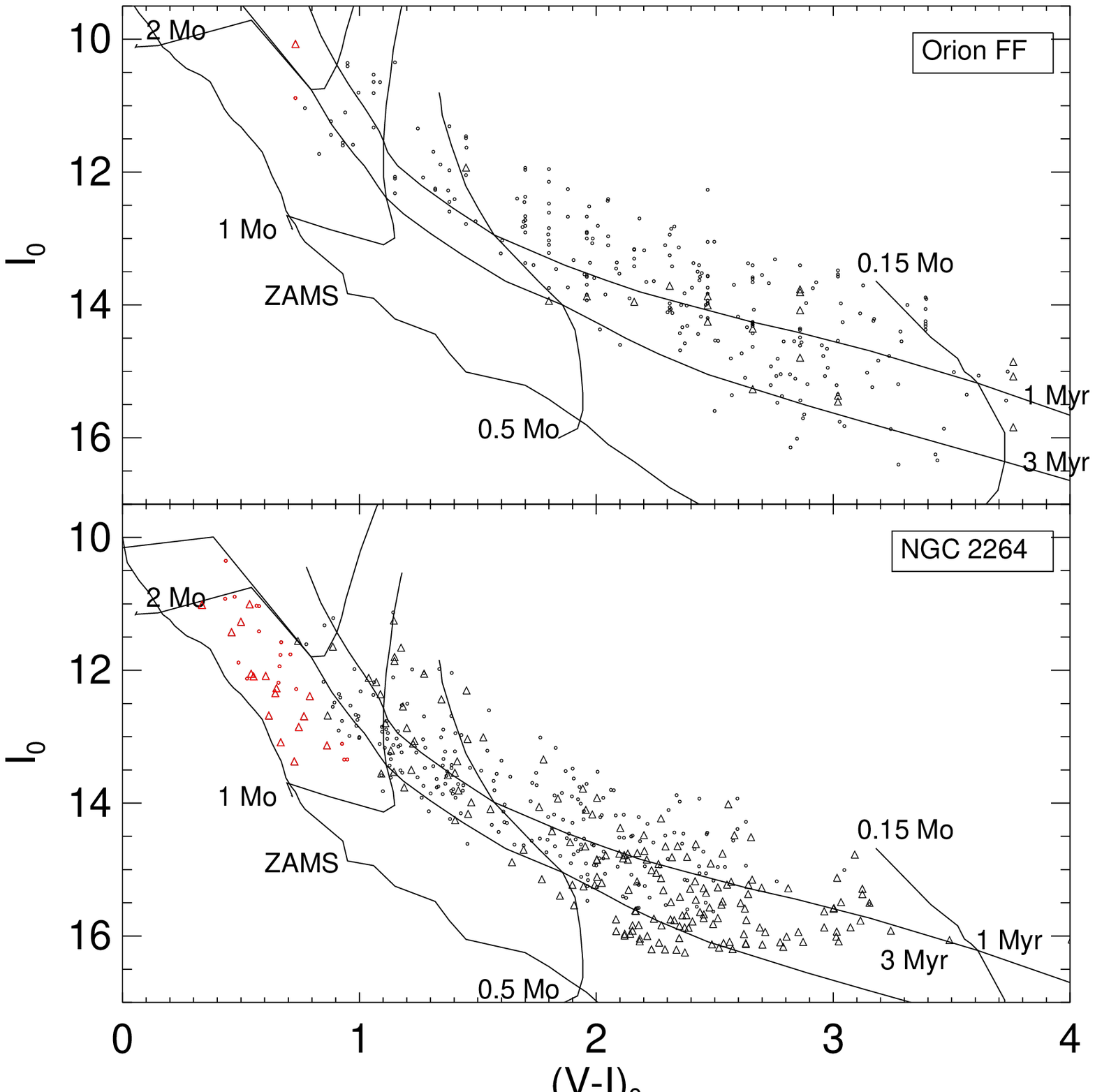}{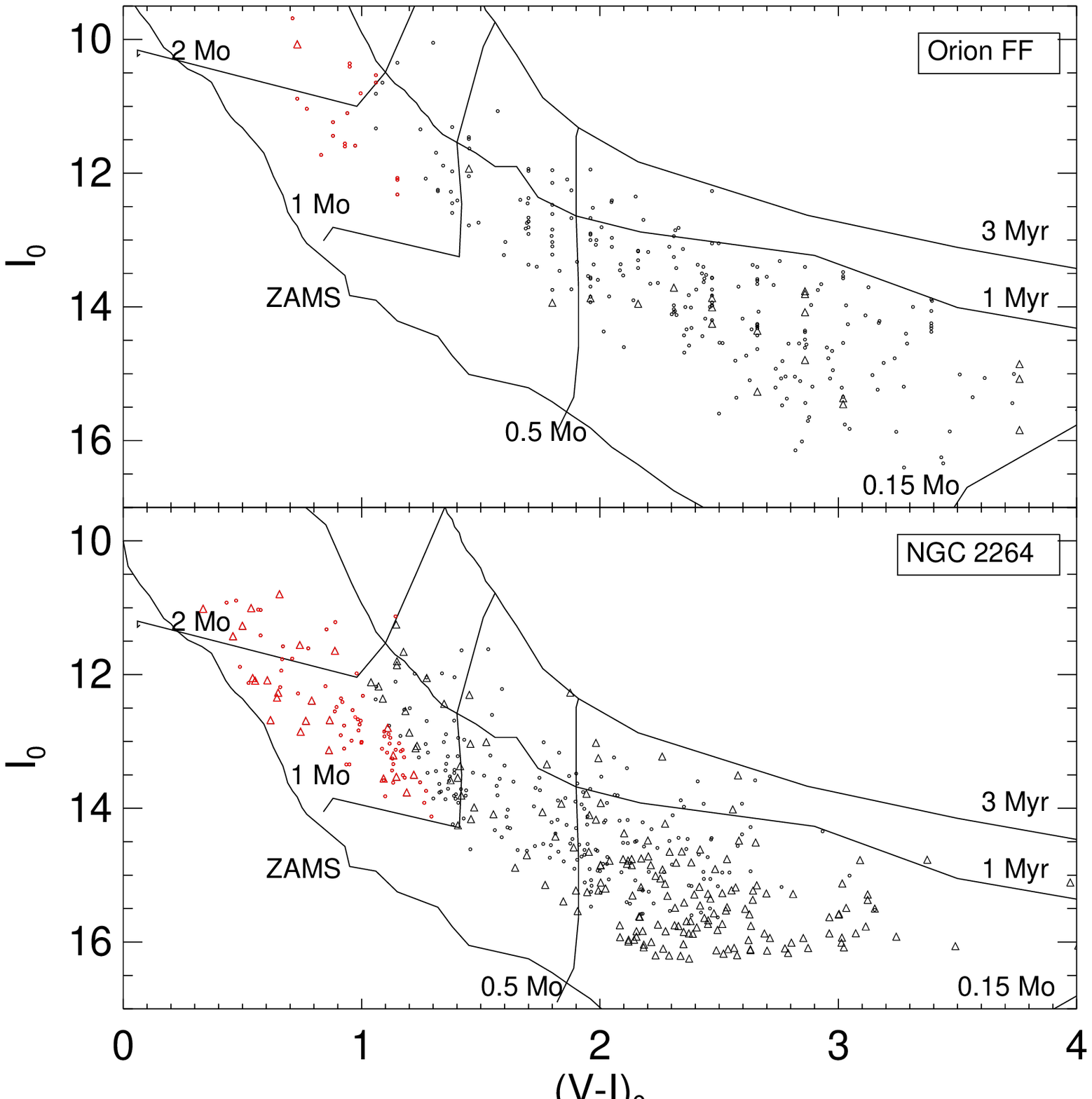}
\caption{Color-magnitude diagram (CMD) for the final best sample to
be considered (see text).  Dots indicate \lx\ detections, and
triangles indicate \lx\ upper limits.  Isochrone lines overplotted
are, from top to bottom, isochrones for 1 Myr and 3 Myr, and the
ZAMS; the left panel is DAM98 and the right panel is SDF.  Lines of
constant mass are, left to right, 2, 1, 0.5, and 0.15 \msun. The
apparent quantization arises from correcting the V-I colors to those
expected from the spectral type. The {\em range} of ages found in
these clusters is $\sim$3 Myr, but the largest {\em fraction} of older
stars is found in NGC 2264. The red symbols denote those stars that
are no longer on convective tracks; see discussion in
\S\ref{sec:oldmass} below.}
\label{fig:cmd}
\end{figure*}


In order to interpret our X-ray data, we have to include data
from the ancillary data catalog (ADC).  However, we cannot
simply take all of the ADC data merged with the X-ray data and
proceed.   Not all stars have a complete set of data in the ADC,
and there is a wide range of ADC survey depths among our sample
clusters.  In this section, we describe what kinds of data we
require for a given star to appear in subsequent analysis in
this paper.  In the end, we settle on a sample which we will
hereafter refer to as the ``best possible sample'' (as defined
below and summarized in Table~\ref{tab:bestposssample}).  Within
this best possible sample, there are stars that are on radiative
tracks according to either the DAM or SDF models, and stars that
are on convective tracks according to those models.  These
samples are referred to as the best possible radiative and best
possible convective samples.  In plots that follow, where both
the radiative and convective stars appear, the radiative stars
are colored red; otherwise, the plots contain just the
convective stars. The following discussion is summarized in
Table~\ref{tab:bestposssample}, where the numbers of detections
and/or upper limits are tabulated for each of a variety of
criteria.

While there are 416 X-ray sources detected in the Orion FF and
683 sources detected in NGC 2264 (line A in
Table~\ref{tab:bestposssample}), only 84\% (348) of the FF
sources and 75\% (509) of the NGC 2264 sources have optical
and/or NIR counterparts in the ADC (line B in
Table~\ref{tab:bestposssample}).  For determining, among other
things, masses and ages, we require a $V$ and $I$ magnitude so
that we can place all the stars in the same CMD and at least
derive self-consistent values.  There are 278 (67\%) of the FF
X-ray detections and 465 (68\%) of the NGC 2264 X-ray detections
meeting those criteria (line D in
Table~\ref{tab:bestposssample}).  Further, of those, 94\% (261
in the FF and 438 in NGC 2264) are likely cluster members based
on position in the sky and position in the CMD (line E in
Table~\ref{tab:bestposssample})---these objects appear in a
clear ``locus'' above the ZAMS (see Rebull \etal\ 2000, 2002 for
full CMDs and additional discussion).   

Of the entire set of X-ray detections, only 115 (27\%) Orion FF
sources and 294 (52\%) NGC 2264 sources can be identified with
counterparts that have known spectral types in the ADC (line C
in Table~\ref{tab:bestposssample}).  Because of the original
motivations behind the studies amassing most of these spectral
types, the set of stars with types is strongly biased toward the
stars with known rotational periods.  As we will see below,
stars with measured periods are brighter in X-rays, which
therefore affects the set of stars with both X-ray detections
and spectral types. 

As a result of this bias, we chose {\em not} to limit our
analysis here to solely the stars with spectral types.  In order
to minimize the chances of field star contamination, we have,
however, restricted our discussion to objects (261 detections in
the FF and 438 detections in NGC 2264) that are likely members
based on position in the sky and in the CMD (line E in
Table~\ref{tab:bestposssample}).  

Upper limits can be important for determining trends in complex
data sets such as this.  We obtained upper limits in \lx\ for
stars with optical detections in the ADC.  For the FF, this adds
84 upper limits to the 261 detections to be considered (for just
the stars that are likely to be members -- the ones in the CMD
locus -- an increase of $\sim$30\%).  The optical surveys of NGC
2264 incorporated into the ADC are very deep and are not
well-matched to the relatively shallow Chandra survey.  If the
entire optical sample from the ADC were included (line F in
Table~\ref{tab:bestposssample}), the number of X-ray upper
limits would dwarf the detections (and moreover would likely be
dominated by non-members).  We have, therefore, limited our
analysis to likely member stars (line G in
Table~\ref{tab:bestposssample}; those both on the molecular
cloud and in the CMD locus) with $I<16$, which is well-matched
both to the \lx\ detections and to the optical surveys
incorporated into the ADC for both clusters.  In order to make a
fair comparison with the set of detections, we further limit
even the \lx\ detections to those stars with $I<16$.  This adds
168 upper limits to the now-reduced 349 detections to be
considered in NGC 2264 (line H in
Table~\ref{tab:bestposssample}).  These restrictions should also
reduce the contamination of our sample by non-members.  The
percentage of non-members is probably higher among stars with no
detections (because members are more likely to be bright in
X-rays) and at the faint end (because the mass function means
there are simply more low-mass stars in the field).  This cut in
$I$ of course has implications for the mass and age range to be
considered here; see Figure~\ref{fig:cmd}.  This cut eliminates
stars with $M \lesssim$0.3 \msun\ and $t>$5 Myr, using DAM
models.

To summarize the discussion to this point (line H in
Table~\ref{tab:bestposssample}), then, there are 261 X-ray
detections with optical counterparts from the ADC that are
likely stellar members of the Orion FF, with 84 additional upper
limits.  There are 349 X-ray detections with optical
counterparts from the ADC with $I<16$ that are likely stellar
members of NGC 2264, with 168 additional upper limits. 

Because we will be comparing the X-ray properties of stars on
radiative tracks with the stars on convective tracks, we have
further limited the discussion to just stars that fall within
the range of $L$ and \teff\ spanned by published stellar models
(the ``model grid''), and those with $M<$2.5 \msun.  This leaves
250 detections and 80 upper limits in the FF, and 317 detections
with 139 upper limits in NGC 2264 (lines I and J in 
Table~\ref{tab:bestposssample})

The sample considered here {\em uses the best available
information for each star}, e.g., if there is a spectral type in
the ADC, the star is dereddened by values specific to each star,
and if there is no type available, the star is dereddened by the
most likely reddening in the direction of the cluster (see
Rebull \etal\ 2000, 2002 for more discussion on reddening
towards these clusters).  A CMD for this sample appears in
Figure~\ref{fig:cmd}.  {\em This constitutes the ``best possible
sample'' for both clusters.}  Note that there are no stars
included in this sample that fall outside the model grid.  Note
also that despite using two different models, DAM and SDF, the
number of stars in the best possible sample using DAM models is
very similar to the best possible sample using SDF models.  

For each of the evolutionary models (DAM \& SDF) depicted in
Figure~\ref{fig:cmd}, the red symbols indicate the stars between
$\sim$1 and $\sim$2.5 \msun\ that are on radiative tracks (lines
K and L in Table~\ref{tab:bestposssample}).  There are very few
of these radiative stars available in the FF, with only 1 X-ray
detection (and 1 upper limit) using DAM models, and 18 X-ray
detections (and 1 upper limit) using SDF models.  But, in NGC
2264, using the DAM models, there are 17 X-ray detections and 16
upper limits; using the SDF models, there are 62 X-ray
detections and 24 upper limits .  {\em The red, radiative points
here are the ``best possible radiative sample.''}

Essentially all of the discussion that follows is restricted to
stars on convective tracks with masses $\leq$2 \msun, e.g., the
black points in Figure~\ref{fig:cmd}.  {\em The black,
convective points here are the ``best possible convective
sample.''}  For the best possible convective sample, using DAM
models, this leaves 249 detections and 79 upper limits in Orion
so 24\% of the best possible convective sample is upper limits;
in NGC 2264, there are 300 detections and 123 upper limits, so
29\% of the best possible convective sample in this cluster is
upper limits.   Similar numbers are obtained for SDF models; see
Table~\ref{tab:bestposssample}. In most dicussions that follow,
we consider only the convective sample.  Where the radiative
stars are included in subsequent discussion and plots, we note
this explicitly.  In plots that follow, where both the radiative
and convective stars appear, the radiative stars are colored
red.   

Of this final, best possible sample of convective stars, we also
want to know how many stars have measured periods and \vsini\
values (lines O and P in Table~\ref{tab:bestposssample}). There
are 110 stars in the Orion FF with measured periods that are
also detected in X-rays, and 14 stars with measured periods but
upper limits in X-rays; there are 64 stars with measured \vsini\
and X-ray detections, and 4 with upper limits in X-rays.  In NGC
2264, there are 181 stars with measured periods and X-ray
fluxes, and 38 more with upper limits in \lx.  There are only 30
stars with measured \vsini\ and X-ray detections, and 3 more
stars with upper limits in \lx.   

\begin{deluxetable}{lllll}
\rotate
\tablecaption{Summary of selection criteria used to assemble
``best possible sample.''
\label{tab:bestposssample}}
\tablewidth{0pt}
\tablehead{
\colhead{} & \multicolumn{2}{c}{Orion FF} &
\multicolumn{2}{c}{NGC 2264} \\
\colhead{criterion/a} & \colhead{detections} & 
\colhead{limits} & \colhead{detections} &
\colhead{limits} }
\startdata
A. X-ray detections & 416 & \nodata & 683 & \nodata \\
B. line A + opt/NIR counterpart & 348 (84\% of A) & \nodata &
	509 (75\% of A) & \nodata \\
C. line B + spectral type & 115 (27\% of A) & \nodata &
	294 (52\% of A) & \nodata \\
D. line B + V, I detection & 278 (67\% of A) & \nodata &
	465 (68\% of A) & \nodata \\
E. line D + inferred cluster membership & 261 (94\% of D) & 
	\nodata & 438 (94\% of D) & \nodata \\
F. upper limits for opt/NIR object in catalog & \nodata & 
	147 & \nodata & 1388 \\
G. upper limit for opt/NIR likely member & \nodata & 84 &
	\nodata & 479 \\
H. lines E \& G + $I<$16 (N2264 only) & \nodata & \nodata &
	349 & 168 \\
I. {\bf BEST POSSIBLE SAMPLE} \\
 \hspace{0.5cm} line D (FF), line H (N2264) + DAM $M<$2.5\msun +
	age exists & 250 & 80 & 317 & 139 \\
J. line D (FF), line H (N2264) + SDF $M<$2.5\msun +
	age exists & 258 & 80 & 327 & 146 \\
K. line I, radiative tracks only, DAM models & 1 & 1 & 17 & 16\\
L. line J, radiative tracks only, SDF models & 18 & 1 & 64 & 24\\
M. line I, convective tracks only, DAM models & 249 & 79 & 300 & 123 \\
N. line J, convective tracks only, SDF models & 240 & 79 & 263 & 122\\
O. line M + periods & 110 & 15 & 181 & 38 \\
P. line M + \vsini & 64 & 5 & 30 & 3 \\ 
\enddata
\end{deluxetable}


\section{$L_{\rm x}$ vs.\ $L_{\rm bol}$}
\label{sec:lxlbol}

\begin{figure*}[tbp]
\epsscale{0.5}
\plotone{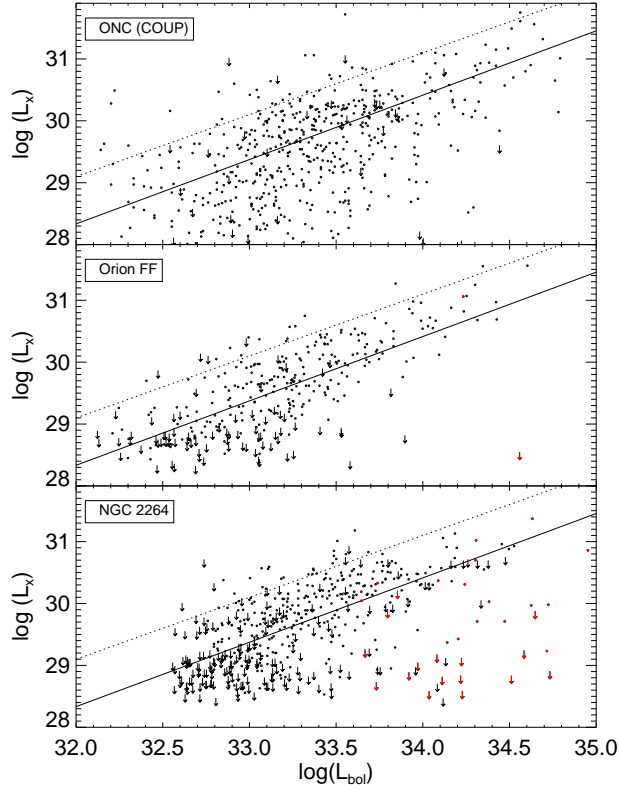}
\caption{Plot of $\log L_x$ vs. $\log L_{\rm bol}$ for the stars
(see \S\ref{sec:bestsample}) in the Orion ONC from the COUP
sample (top plot), Orion FF (middle plot) and NGC 2264 (bottom
plot).  Only detections from COUP are shown, whereas upper
limits are shown for the other two panels.  The upper limits
occur at all values of \lx\ because the PSF varies significantly
across the field of view, so a star appearing at the edge of the
field will have a much higher upper limit than a star near the
center of the field.  The boundaries for this plot were
optimized for the FF and NGC 2264 data; 24 detections and 14
limits from the COUP data appear outside of this region, all but
4 detections (and 1 limit) of which appear below log \lx=28. The
solid line is the best fit value to the ONC COUP data analyzed
by Preibisch \etal\ 2005b. The dotted line is the canonical
``saturation'' value of $\log L_x/L_{\rm bol} \sim -2.9$.  In
the lower two panels, we see a suggestion of a bifurcation in
the distribution beginning at about half the luminosity of the
Sun and extending to higher luminosities.  In NGC 2264, 50\% of
the stars on the lower branch are on radiative tracks (indicated
in red) and therefore may have X-ray properties that are
different from stars on convective tracks.   A discussion of the
comparison with the ONC is provided in the text.}
\label{fig:lxlbol}
\end{figure*}

\begin{figure*}[tbp]
\epsscale{0.5}
\plotone{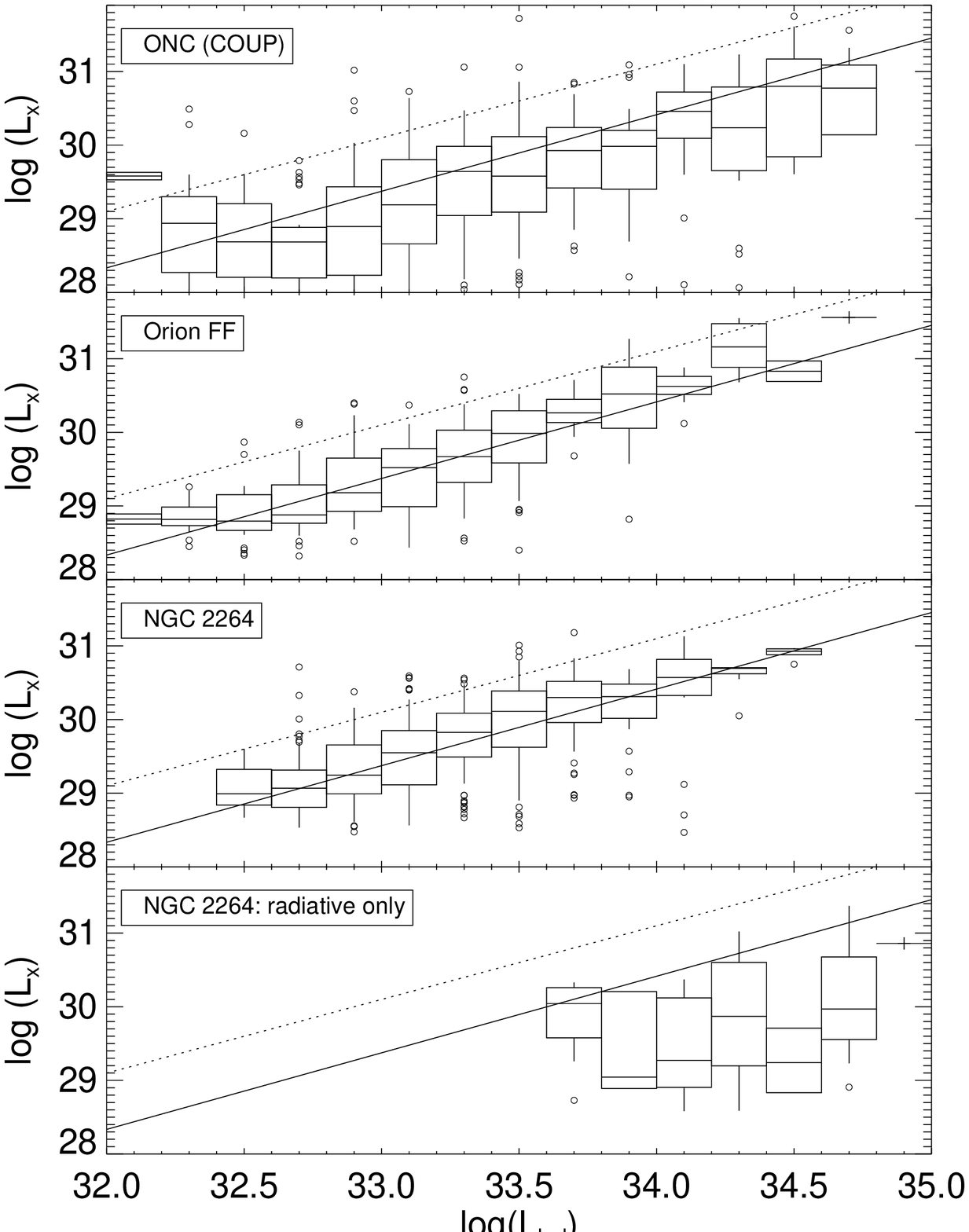}
\caption{Plot of $\log L_x$ vs. $\log L_{\rm bol}$ for the stars
from Figure~\ref{fig:lxlbol}, but represented in box plot
form.  The first three panels are convective stars; the last
panel is the radiative stars in NGC 2264 (there are not
enough radiative stars in the other clusters to merit a separate
plot).  It is readily apparent here that the X-ray properties of
the radiative stars are different than those of the convective
stars. }
\label{fig:lxlbolprime}
\end{figure*}

All recent investigations, including the COUP study  (e.g.,
Preibisch \etal\ 2005b), have found a clear positive correlation
between log \lx\ and log \lbol\, but with a very large intrinsic
(as opposed to instrumental) scatter (Feigelson \etal\ 2003).

Figure~\ref{fig:lxlbol} shows the results for the ONC from the
COUP sample, Orion FF, and NGC 2264.   The best possible
radiative sample for NGC 2264 and the Orion FF is shown in red,
and the best possible convective sample is shown in black
(essentially all of the COUP stars are convective).  Also shown
is the best fit straight line obtained by Preibisch \etal\
(2005b) for the ONC COUP data, which has a sensitivity limit of
about log \lx=27.0.  This should be compared with a sensitivity
limit of about log \lx\ $\sim28.5$ for our observations. The
line derived for the ONC is a reasonable fit to the data for
stars on convective tracks for both of our regions.  We note,
however, that many of our points near log \lbol$\sim$33.5 lie
above the mean line for the ONC, and there is a cloud of points
with low values of log \lx\ that lie below the line near log
\lbol$\sim$32.8, suggesting that perhaps a linear fit may not be
appropriate.  However, both of these features are also seen in
the ONC plots and may thus reflect some inadequacy in the
ability of a straight line to fit the data for both data sets. 
Figure~\ref{fig:lxlbolprime} depicts the same data as in
Figure~\ref{fig:lxlbol}, but in box plot form.  These box plots
have been used in other papers (e.g., Flaccomio \etal\ 2003c) as
a mechanism for interpreting scatter plots.  We emphasize that
the boxes in Fig.~\ref{fig:lxlbolprime} have been corrected
using the Kaplan-Meier estimator for censored data to take into
account the upper limits present in the data.  The central line
in each box denotes the (corrected) median; the ends of each box
are the first and third quartile of the (corrected)
distribution; the lines extend to the most extreme values that
are not more than 1.5 times the interquartile range; and the
open circles are those points outside 1.5 times the
interquartile range.  If downward-pointing triangles appear in
the boxes, then the true (corrected) lower limit of the box is
located at some unknown location below the line indicated.  The
top three panels in Fig.~\ref{fig:lxlbolprime} contain only
convective stars; the bottom panel contains only the radiative
stars in NGC 2264 (there are not enough radiative stars in Orion
to merit a separate plot).  Examination of
Fig.~\ref{fig:lxlbolprime} reveals that the X-ray properties of
the radiative stars are significantly different than those of
the convective stars.

Preibisch \& Feigelson (2005) used the Orion data to search for
changes in X-ray emission with age.  They assumed that the stars
in Orion ranged in age from 0.1 to 10 Myr (Palla \& Stahler
1999).  Because contraction down the Hayashi tracks slows with
increasing age, at ages of $\sim$3 Myr, typical uncertainties in
derived luminosities and effective temperatures lead to age
uncertainties of $\sim$2-3 Myr (1$\sigma$ errors, Hartmann
2001).   Based on the ONC data alone,  Preibisch \& Feigelson
concluded that the mean \lx\ decayed slowly with age (\lx\
$\propto\tau^{-1/3}$) during the first 10 Myr for stars with
masses in the range 0.5-1.2 \msun.  However, in order to match
the X-ray properties of ZAMS stars in young clusters, the decay
of \lx\ must be accelerated at ages $t>$10 Myr.  For stars with
masses in the range 0.1-0.4 \msun\ they found only a modest
decrease in log \lx\ during the first 100 Myr and a possible
slight increase in log \lx/\lbol\ over that same time period.
Flaccomio \etal\ (2003c) also investigate trends of X-ray
emission with time, comparing median levels of X-ray emission in
several star-forming regions.  They found constant log \lx\ and
an increase in log \lx/\lbol\ with time through $\sim$3-4 Myr and
a subsequent leveling off at saturation level (log \lx/\lbol\ 
$\sim -3$).

Our own data are for clusters which contain stars that are on
average older than the stars in the ONC.  We see two differences
with respect to the ONC data. First, the upper envelope in log
\lx\ for ONC data lies about 1.5 dex or a factor  of 30 above
the best fit straight line.  For the Orion FF and NGC 2264, the
upper envelope is only about a factor of 10 (1 dex) above the
best fit straight line. As Fig.~\ref{fig:lxlbol} also shows,
very few of the Orion FF and NGC 2264 stars lie above the
nominal ``saturation'' limit of log \lx/\lbol\ $= -2.9$, while a
large number of ONC stars lie above this limit (Preibisch \etal\
2005b). Quantitatively, 14\% of the detected ONC stars have  log
\lx/\lbol\ $\geq -2.9$,  and 4.6\% have log \lx/\lbol\ $\geq
-2.5$.  For the Orion FF, the fractions are 5\% and 0.4\%,
respectively, and for NGC 2264, the fractions are 7\% and 0.9\%
respectively.  In other words, there is a significantly larger
number of stars with high \lx\ and high \lx/\lbol\ in the
younger ONC.  There are three possible explanations for this
effect.  First, it could be entirely an age effect, in that the
most {\em extreme} levels of activity may decay during early PMS
evolution.  Second, it could be the intrinsic variability of the
stars themselves in \lx.  Third it could be the intrinsic
variability of the stars themselves in the optical photometry
that goes into the \lbol\ estimate.  In either of these latter
cases, the variability could be more extreme in the younger
stars, allowing for more outliers.  With regards to the
variability in \lx, Favata \etal\ (2005), in studying the
brightest $\sim$1\% of the COUP flares, find some objects that
flare 1-2 orders of magnitude, and that the characteristic times
for these flares range from 10-400 ks.  In the specific case of
the COUP sample, however, as Preibisch \& Feigelson (2005) point
out, the influence of individual flares on the total \lx\
reported for any one star is reduced by the long time baseline
of their observations.  Thus, outliers solely due to flares in
the COUP data set are less likely than in the other (much
shorter) observations.  Variability in the optical data that go
into calculation of \lbol\ may account for some fraction of the
outliers; optical variability can exceed a magnitude.   Rebull
\etal\ (2002b) discussed the sources of these errors in detail
for stars with spectral types over limited mass ranges and in
regions of less reddening.  For the COUP sample in particular,
even though reddening is a concern, spectral types are known for
every object, and so it is unlikely that this can account for
all (or even most) of the outliers.  More data on time
variability will be required to determine why there are more
outliers with high \lx\ in the ONC.

The lower envelope of the X-ray detected stars in all three
regions for both \lx\ and \lx/\lbol, however, is comparable for
stars with \lbol\ similar to that of the Sun, and so the decay
in the {\em mean} and {\em lower bounds} of the X-ray
luminosities with time during the first 5 Myr of PMS evolution
indeed appears to be small, as Preibisch \& Feigelson (2005)
reported for the ONC alone.  Unfortunately, because of the
shallow depth of our current survey, we cannot compare the lower
bounds of the detections of stars with M $<1$ \msun\ in these
three regions.  Since more than half the stars in our sample
were detected, we can compare the {\em median} log \lx\ and log
\lx/\lbol\ for the best possible convective sample in all three
clusters.  The median log \lx\ for the ONC (further restricted
to $M>$0.16 \msun to account for the difference in survey
depths), the Orion FF, and  NGC 2264 are 29.55, 29.71, and
29.90, respectively; the median values of log \lx/\lbol\ (in the
same order) are $-3.71$, $-3.50$, and $-3.44$.  (There are some
upper limits in this sample, so as a simple worst-case test, we
recomputed the medians, adopting the upper limits for all stars
as detections, and it does not change these median values very
much.)  These results are consistent with that obtained by COUP
for the ONC.  While it is true that the most luminous objects
have disappeared by the age of NGC 2264, there is not much
change in the median values. We certainly do not see a
significant decrease during the first 5 Myr, in agreement with
the Preibisch \& Feigelson (2005) work on the ONC alone.  

\section{X-ray properties and stellar structure: the transition from
convective to radiative tracks}
\label{sec:oldmass}


\begin{figure*}[tbp]
\epsscale{0.5}
\plotone{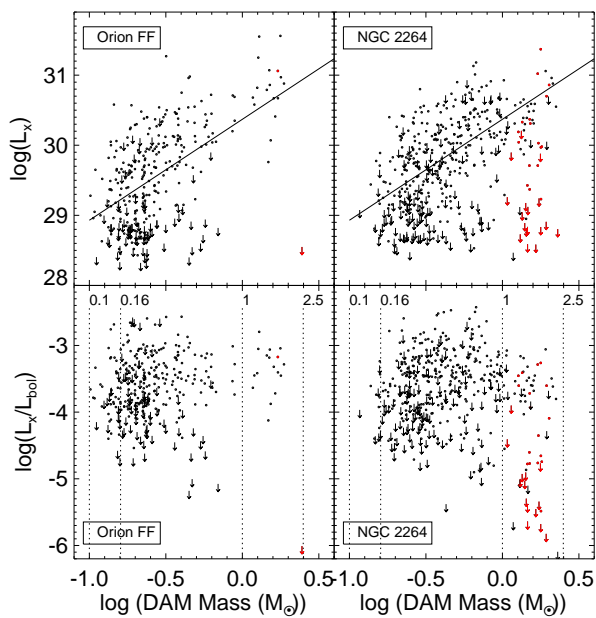}
\caption{Plot of $\log L_x$ (top plots) and $\log L_x/L_{\rm bol}$
(bottom plots) vs. $\log M/M_{\sun}$ from DAM for the Orion FF (left
plots) and NGC 2264 (right plots), for the best samples available for
each cluster (see \S\ref{sec:bestsample}).  The solid line is the
relationship between \lx\ and SDF masses found for the ONC by
Preibisch \etal\ 2005b. Note again the apparent bifurcation at about
0.5 \msun. Red symbols are those stars still on radiative tracks; see
Fig.~\ref{fig:cmd}.  Dotted lines indicate 0.1, 0.16, 1, and 2.5 
\msun, from left to right.}
\label{fig:mass}
\end{figure*}


\begin{figure*}[tbp]
\epsscale{0.5}
\plotone{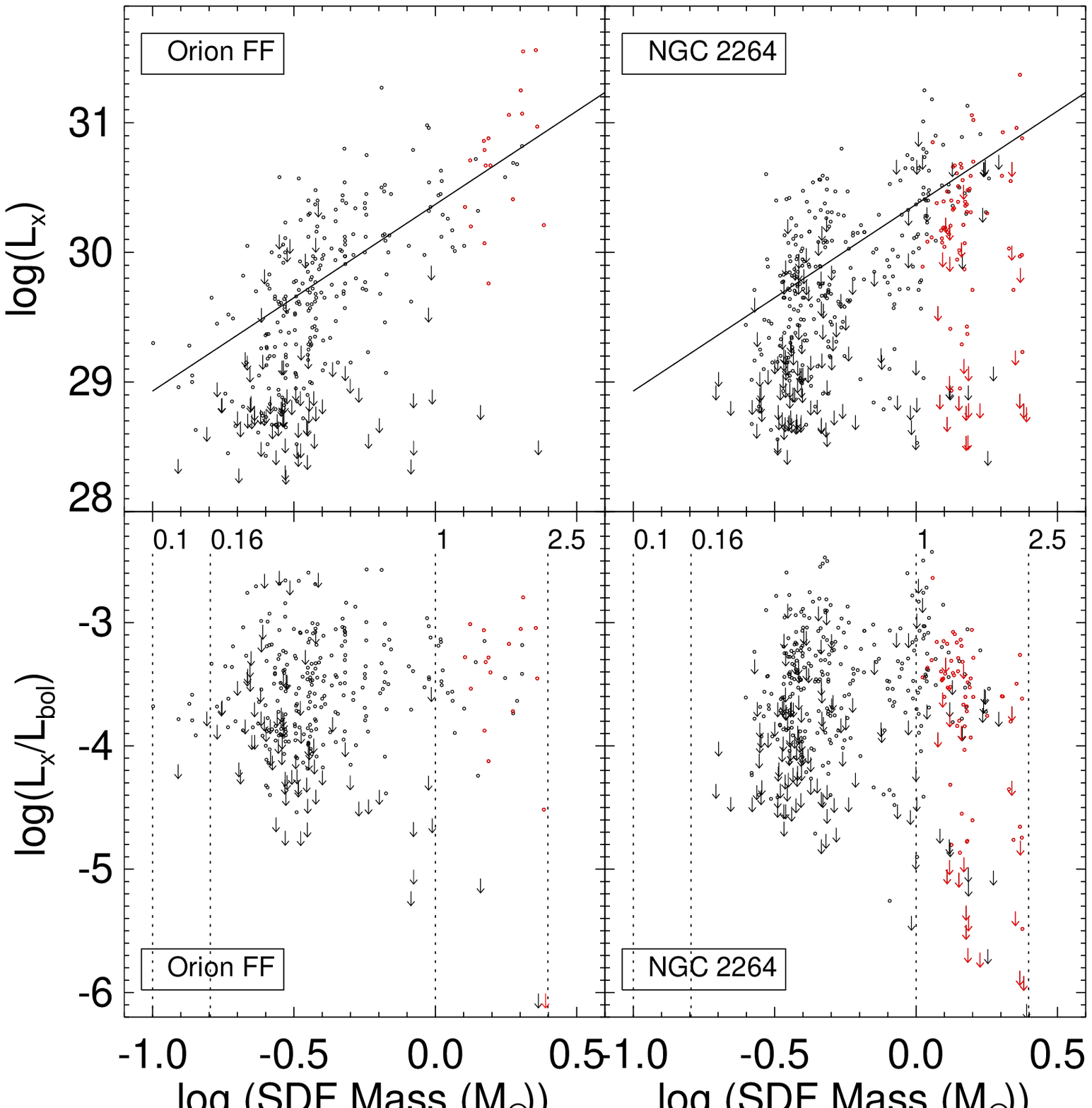}
\caption{Plot of $\log L_x$ (top plots) and $\log L_x/L_{\rm bol}$
(bottom plots) vs. $\log M/M_{\sun}$ from SDF for the Orion FF (left
plots) and NGC 2264 (right plots), for the best samples available for
each cluster (see \S\ref{sec:bestsample}).  The solid line is the
relationship between \lx\ and SDF masses found for the ONC by
Preibisch \etal\ (2005). Note again the apparent bifurcation at
about 0.5 \msun. Red symbols are those stars still on radiative
tracks; see Fig.~\ref{fig:cmd}.  Dotted lines indicate 0.1, 0.16,
1, and 2.5 \msun, from left to right.}
\label{fig:masssdf}
\end{figure*}


In this section, we compare the X-ray properties of low mass
stars still located on fully convective tracks with those of
higher mass stars, some of which appear to lie on convective
tracks, while others of slightly greater age have developed
radiative cores and are proceeding toward the main sequence
along radiative tracks. In Figure~\ref{fig:cmd}, the stars
on radiative tracks are indicated.

In Figures~\ref{fig:mass} and \ref{fig:masssdf}, we plot log
\lx\  and log \lx/\lbol\ vs.\ mass, retaining the red
(radiative) and black (convective) convention. Among the
detections, the stars on radiative tracks have lower values than
their similar mass counterparts located on convective tracks;
moreover, there is a higher fraction of upper limits among the
radiative stars than among the convective stars.  Examining
solely stars with detections in NGC 2264, the median log
\lx/\lbol\ for all convective stars between 1 and 2.5 \msun\ is
$-$3.51; the median log \lx/\lbol\ for radiative stars (over the
same mass range) is $-$4.01.  However, many of the values for
radiative stars are upper limits; adopting the upper limits for
all stars as detections at the level of their upper limits, the
median log \lx/\lbol\ for the convective stars changes subtly to
$-$3.54, while the median for the radiative stars changes to
$-$4.48. By assigning the upper limits to be ``detections,'' we
of course obtain a conservative limit on the median log
\lx/\lbol; even so, the medians for the convective and radiative
samples differ by a factor of 10. Evidently the transition from
a fully convective state to a radiative-core convective envelope
configuration results in a dramatic decrease in coronal activity
as diagnosed by X-rays; Gagne \etal\ (1995) also report
differences in the X-ray properties between the convective and
radiative stars in Orion based on ROSAT observations (see also
Strom 1994 and Flaccomio \etal\ 2003b).

We have examined the COUP data to search for a similar trend
among the ONC sample. Unfortunately, because the region is
younger, too few stars have evolved to their radiative phases to
enable a convincing test.

Preibisch \etal\  (2005b) reported a weak but significant
correlation between log \lx/\lbol\ and mass.  Spearman's
correlation coefficient does not find any significant
correlations in our data for the Orion FF and NGC 2264. However,
we note that our survey covers a much smaller mass range than
COUP.  Restricting the comparison with COUP data to the same
mass range reveals no clear trends in either sample.  Because
our data are dominated at low masses by upper limits, we are
also unable to search for mass-dependent differences in log
\lx/\lbol\ as reported by Flaccomio \etal\ (2003b, 2003c), who
find similar log \lx\ vs.\ mass relationships to that reported
by Preibisch \etal\  (2005b). 

The conclusion that X-ray luminosity is related to the interior
structure of PMS stars is of course valid if and only if stars
assigned to radiative tracks are placed correctly in the HR diagram,
are predominantly members of the Orion FF and the ONC, and if the
X-ray flux is not dominated by a faint companion that is not visible
in the optical.  We now examine each of these requirements for  NGC
2264, which has the largest number of massive stars in apparent
post-convective evolution. 

How accurately have we located these stars in the HR diagram?  These
stars are among the brightest stars in the cluster, and for 94\% of
the radiative stars (both detections and upper limits), we have
spectral types; they range in type from A0 to K4.  Because we have
types, we are able to deredden each star individually with a
correction appropriate to that specific star.  Therefore, the
placement of these stars in the dereddened CMD would appear to be as
secure as possible for young (often variable) stars.  These stars
appear in the correct vertical position to be taken as cluster
members; they do not appear significantly above or below the ZAMS for
a cluster at the distance of NGC 2264.  We therefore believe that
most of these stars are members based on position in the HRD.

Because these stars are among the brighter stars in the cluster,
we have preliminary proper motions (see Rebull \etal\ 2002) for
54\% of the stars on radiative tracks; 33\% of the red points
(detections and  upper limits) in Figure~\ref{fig:cmd} (DAM
models) are likely members, 22\% are likely non-members, and the
rest have no proper motion membership information available.  
Of the available sample, at least 60\% of the stars are thus 
members based on proper motions. 

Another indicator of likely membership is the presence of a
circumstellar disk.  Seventy percent of the best possible radiative
sample has positive circumstellar and/or accretion disk indicators,
suggesting at least 70\% of the sample are indeed members.   

Are unresolved binaries important?  They may play a role,
particularly in the case of X-rays where low-mass companions can
outshine their primaries.  However, what we have shown is that
the fluxes of the stars on radiative tracks are systematically
{\em lower} than those of stars with similar masses on
convective tracks.  If optically fainter companions do
contribute some of the flux, then the discrepancy must be even
larger.

In summary, then, we believe that most ($>$70\%) of the best
possible radiative sample are indeed cluster members.  We
therefore conclude that there is a significant decrease in X-ray
luminosity as stars of similar mass evolve from convective
tracks, where log \lx/\lbol\ values lie close to the
``saturation limit," to radiative tracks, where \lx/\lbol\
values are $\sim$10 times smaller.  In support of this result we
note two additional observations:  1) the apparent `bifurcation'
in \lx/\lbol\ disappears for masses lower than about 0.5 \msun\,
which is approximately the mass below which stars remain fully
convective from the stellar birthline to the ZAMS; and 2) the
levels of \lx\ and \lx/\lbol\ for stars on radiative tracks are
roughly in agreement with the values found in this mass range in
the NEXXUS survey of nearby field stars (Schmitt \& Liefke
2004).  

\clearpage

\thispagestyle{empty}

\begin{deluxetable}{lllllll}
\rotate
\tabletypesize{\tiny}
\tablecaption{Summary of recent studies of large surveys of star-forming
regions considering effects of disks/accretion/degree of
embeddedness or period on $L_{\rm x}$
\label{tab:diskliterature}}
\tablewidth{0pt}
\tablehead{
\colhead{work} & \colhead{instrument} & \colhead{cluster} & \colhead{range
of $L_x/L_{bol}$\tablenotemark{a}} &\colhead{disk indicator} &
\colhead{conclusion on disks\tablenotemark{b}} &
\colhead{conclusion on rotation}}
\startdata
Bouvier (1990) & Einstein & Taurus-Auriga &  log\lx=27-32 &
	H$\alpha>$10\AA & no diff.\ & anti-correl. \\ 
Feigelson \etal\ (1993) & ROSAT/PSPC & Cham I & $-$2 to $-$5 &
	EW(H$\alpha$)$>$10\AA &  $L_{\rm x}$(C)$<$$L_{\rm x}$(W) but no
	diff.\ within sel.\ eff.\  & no correl. \\ 
Gagne \& Caillault (1994) & Einstein & ONC & $-$2 to $-$7 
	& $H-K$, H$\alpha$ & no difference & no correl. \\
	& & & ($-$2 to $-$4.5) & & \\ 
Damiani \& Micela (1995) & Einstein & Taurus-Auriga & $-$3 to $-$5 & 
	IRAS-25 & WTTS fainter & weak anti-correl.\ with scatter \\
Gagne \etal\ (1995) & ROSAT/HRI & ONC & $-$2 to $-$7  & 
	$\Delta(H-K)>0.1$ &  $L_x, L_x/L_{bol}$(C)$<$$L_x, L_x/L_{bol}$(W)
	& no correl. \\
	& & & ($-$2 to $-$5) & & \\ 
Casanova \etal\ (1995) & ROSAT/PSPC & $\rho$ Oph & $-$2 to $-$5 &  
	Class 0-3 & $L_{\rm x}$(embedded)$\sim$$L_{\rm x}$(less embedded), no
	diff. & N/A \\ 
Neuhauser \etal\ (1995) & RASS & Taurus-Auriga & $-$4 to $-$7 & 
	EW(H$\alpha$)$>5$-15 \AA &  $L_{\rm x}$(C)$<$$L_{\rm x}$(W) & 
	correl. \\
	 & & & & &  $L_{\rm x}$(C) harder than W\\ 
Lawson \etal\ (1996) & ROSAT/PSPC & Cham I & $-$2 to $-$5 & H$\alpha$ & no
	difference & N/A \\ 
Preibisch (1997) & ROSAT/PSPC & several &  log\lx=27-32 & N/A & N/A &
	anti-correl.\\
Alcala \etal\ (2000) & RASS & Orion &  log\lx$\sim$31 & N/A & N/A & no
	correl.\ for K stars\\
Wichmann \etal\ (2000) & RASS & Taurus-Auriga & $-$2.5 to $-$5 & N/A & N/A
	& correl. \\
Flaccomio \etal\ (2000) & ROSAT/HRI & NGC 2264 & $-$2 to $-$5 
	 & H$\alpha$ & no difference, but C more variable & N/A \\
	 & & & ($-$2 to $-$4) & & \\ 
Grosso \etal\ (2000) & ROSAT/HRI & $\rho$ Oph & $-$2 to $-$4 &  ISO data &
	no difference & N/A \\ 
Stelzer \etal\ (2000) &  ROSAT/PSPC & Taurus-Auriga &  & HBC listing W/C
	& $L_{\rm x}$(C)$<$$L_{\rm x}$(W) but stronger flares, &
	anti-correl. \\ 
	& & & & & \& C more variable &  \\ 
Imanishi \etal\ (2001) & Chandra/ACIS & $\rho$ Oph & &  Class 0-3 & 
	$L_x/L_{bol}$(Class I)$<$$L_x/L_{bol}$(Class II \& III) & N/A\\ 
Stelzer \& Neuhauser (2001) & ROSAT/PSPC & Taurus-Auriga & $-$3 to $-$6 & 
	EW(H$\alpha$)$>$ 10\AA & $L_{\rm x}$(C)$<$$L_{\rm x}$(W) & 
	correl. \\ 
Preibisch \& Zinnecker (2002) & Chandra/ACIS & IC 348 & $-$2 to $-$6
	&  EW(H$\alpha$)$ \geq$ 10 \AA, 
	&  $L_{\rm x}$(C)$<$$L_{\rm x}$(W), but likely to be sel.\ eff. &
	N/A\\
	& & & ($-$2 to $-$4.5) & $\Delta(K-L)>0.17$  & (goes away for
	$\Delta(K-L)$ \\ 
Feigelson \etal\ (2002, 2003) & Chandra/ACIS & Orion (ONC) & $-$2.5 to $-$6 &
	$\Delta(I-K)>0.3$ & no difference & weak correl.  \\ 
Getman \etal\ (2002) & Chandra/ACIS & NGC 1333 & &  $JHK$ & no difference &
	N/A \\ 
Tsujimoto \etal\ (2002) & Chandra/ACIS & OMC-2, OMC-3 &  &  
	Class 0-3 & $kT$ of C higher than W, but poss.\ sel.\ eff.
	& N/A\\ 
Flaccomio \etal\ (2003a) & ROSAT/HRI & NGC 2264, & $-$2 to $-$7.8
	&  $\Delta(I-K)>0.8$  & $L_x,
	L_x/L_{bol}$(C)$<$$L_x, L_x/L_{bol}$(W) & N/A \\
	& & Cham I & ($-$2 to $-$5) & & \\ 
Flaccomio \etal\ (2003b) & Chandra/HRC & ONC & $-$2 to $-$7.8 
	& EW(Ca II)$>$ 1 \AA &  $L_x, L_x/L_{bol}$(C)$<$$L_x,
	L_x/L_{bol}$(W) & no correl.; stars with $P$ high \lx \\ 
	& & & ($-$2 to $-$5)& &  \\ 
Stassun \etal\ (2004) & Chandra/ACIS & ONC & $-$2 to $-$6 & EW(Ca II)$>$ 1
	\AA & $L_x, L_x/L_{bol}$(C)$<$$L_x,L_x/L_{bol}$(W) & log \lx/\lbol\
	correl.\ with P, but not  \\
	 & & & & & & \vsini; stars with $P$ high \lx \\ 
Feigelson \& Lawson (2004) & Chandra/ACIS & Cham I & $L_x$=28-31 & various
	& $L_x$(C)$<$$L_x$(W) (slightly) & N/A \\
Preibisch (2005) & XMM/EPIC & Serpens & $\sim -3$ & literature & Class I
	more variable than Class II+III & N/A \\
Ozawa et al.\ (2005) & XMM/EPIC & $\rho$ Oph & $-$2 to $-$5 & literature &
	Class I higher T, absorp.\ than II+III & N/A \\
Preibisch et al.\ (2005b) & Chandra/ACIS & ONC (COUP) & $-1$ to $-$6 & 
	$\Delta(I-K)$,  & accretors less active & weak correl. \\
	& & & & $\Delta(K-L)$, EW(Ca II) & & \\
\enddata
\tablenotetext{a}{Values in parentheses are for the low-mass stars in the
sample.}
\tablenotetext{b}{``C'' = CTTS; ``W'' = WTTS}
\end{deluxetable}

\clearpage

\section{X-ray Emission and Rotation}
\label{sec:rotation}


\begin{figure*}[tbp]
\epsscale{0.5}
\plotone{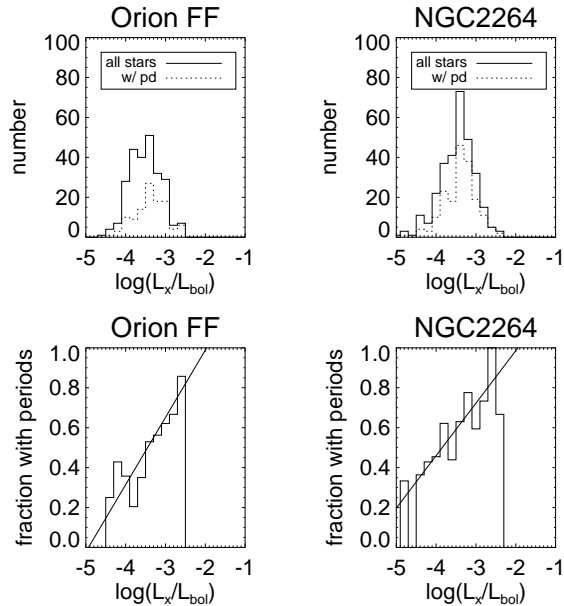}
\caption{The log \lx/\lbol\ distribution for the best samples available
for each cluster (see \S\ref{sec:bestsample}, including only
convective stars) for stars with and without measured periods.  
Stars with periods are X-ray bright, and stars that are X-ray bright
are more likely to have measured periods. The line at an angle is a
fit to the ratio.  The slopes are 0.34$\pm$0.05 (Orion FF) and
0.26$\pm$0.05 (NGC 2264). }
\label{fig:pbiaslxlbol}
\end{figure*}


\begin{figure*}[tbp]
\epsscale{0.5}
\plotone{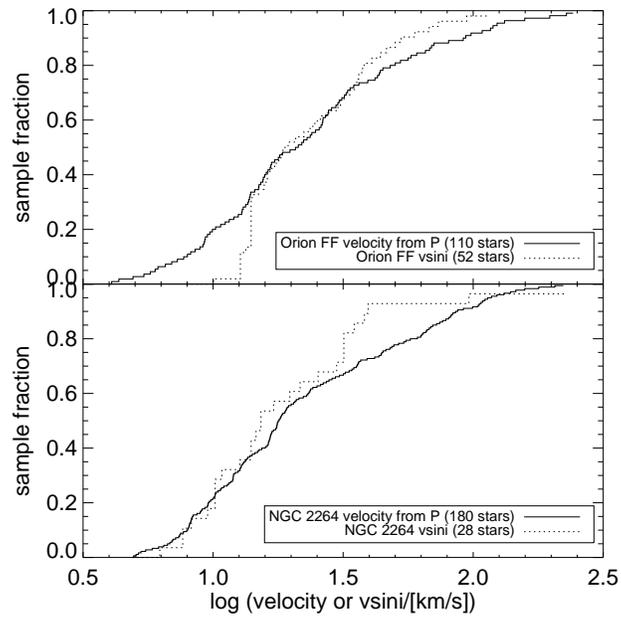}
\caption{Cumulative distributions of measured \vsini\ (multiplied by
4/$\pi$ to account for projection effects; dotted line) and
calculated velocity (solid line) for the FF (top panel) and NGC 2264
(bottom panel).  Aside from where the spectral resolution cuts off the
\vsini\ measurements, the distributions are indistinguishable. }
\label{fig:cumudistv}
\end{figure*}


Flaccomio \etal\ (2003b) first found evidence that stars in the
ONC for which rotation periods are known to have higher X-ray
luminosities than stars which show no clear evidence of
spot-modulated rotation; Stassun \etal\ (2004) confirm this
result.  

The same trend is seen among fully convective stars ($M <$ 0.5
\msun) in both the Orion FF and NGC 2264. 
Figure~\ref{fig:pbiaslxlbol} shows that the stars with derived
periods in these two associations are clearly brighter  in log
\lx/\lbol, and, conversely, stars that are bright in log
\lx/\lbol\ are more likely to have a measurable period.  For the
best possible convective sample (plotted in
Figure~\ref{fig:pbiaslxlbol}), the best-fit slopes to the lower
panels in the Figure are (for the FF) 0.34$\pm$0.05, and (for
NGC 2264) 0.26$\pm$0.05.  These values do not change
significantly when considering other possible subsamples of the
data.  Median values of log \lx/\lbol\ for the best possible
convective sample for the ONC, FF, and NGC 2264, for the sample
with periods, are (respectively) $-3.57$, $-3.35$, and $-3.37$. 
For the best possible convective sample without periods (in the
same order), the medians are $-3.81$, $-3.65$, and $-3.53$.  It
is clear that the samples with measured $P$ are brighter in
X-rays (and again there are no trends with age).

In searching for correlations between X-ray emission and
rotation rates of PMS stars, another concern is whether the
sample of stars with known periods represents the full range of
rotation rates for PMS stars.  It might be, for example, that
periodic stars are biased toward PMS stars dominated by either
long or short rotation periods, or by stars viewed more nearly
equator-on than pole-on.  The study of periods and \vsini\ in
the ONC by Rhode \etal\ (2001) concluded that there was no
difference in the distribution of rotation rates for samples of
stars with periods and samples of stars with \vsini.  A similar
test is performed for the current data set in
Figure~\ref{fig:cumudistv}. For the purpose of this graph,
rotational velocities have been derived from $P$ and an 
estimate of the radius (see, e.g., Rebull \etal\ 2002 for more
on this calculation); the \vsini\ values have been converted to
velocity by multiplying the observed rates by 4/$\pi$ to allow
for inclination effects (Chandrasekhar \& Muench 1950; Gaige
1993).  The Orion FF appears to exhibit a relative paucity of
objects with low \vsini.  However, this reflects the resolution
limit of the data rather than a real physical effect.  Taking
this into account, we conclude from these data that there is no
apparent bias in the period-selected sample.  However, our data
set is too sparse to distinguish whether the \vsini\
distribution for the low \lx/\lbol\ stars that lack periods
differ from the ensemble distribution. 

Why are we more likely to be able to determine periods from
photometric monitoring programs for stars with large values of
log \lx/\lbol? In order to obtain an optical  photometric
period, we  require a clear view of the stellar photosphere
where there are large, stable, relatively isolated spots or spot
groups rotating into and out of view causing periodic
modulations.  Stassun \etal\ (2004)  have suggested that perhaps
more active stars have larger and/or more  organized magnetic
spot coverage and that periodic variations are more easily
measured in such circumstances. If so, then stars with more
organized spots also have higher levels of coronal activity
leading to higher X-ray luminosity.

Since the earliest days of X-ray astronomy, investigators have
sought a relation between rotation rate (periods or \vsini) and
X-ray luminosity for PMS stars.  Because the X-ray flux  is
thought to originate in coronal activity which is driven by
convection and rotation, and because faster-rotating main
sequence stars are brighter in X-rays (e.g., Stauffer \etal\
1994, Micela \etal\ 1996, Pizzolato \etal\ 2003), it was perhaps
reasonable to expect to find a  similar relation in younger
stars, but the evidence remains mixed. 
Table~\ref{tab:diskliterature} contains a listing of the
conclusions on this topic from the literature to date. Most
recently, the  COUP project (Preibisch \etal\ 2005b) found a
weak but statistically significant correlation between P and
log \lx/\lbol.

\begin{figure*}[tbp]
\epsscale{0.5}
\plotone{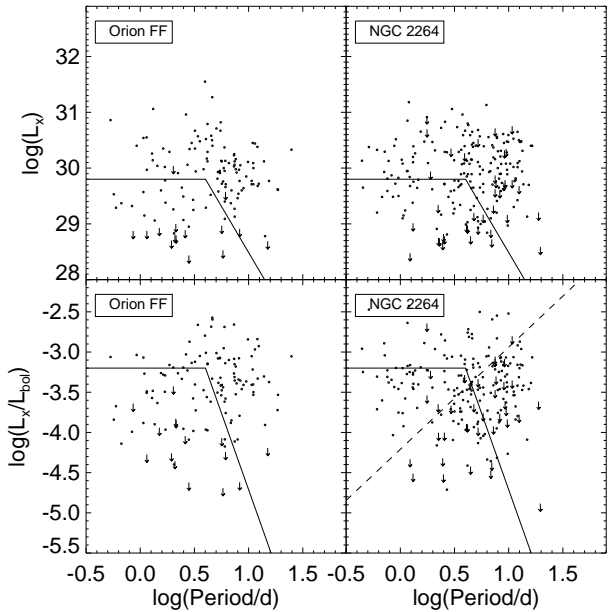}
\caption{Log \lx\ and log \lx/\lbol\ vs.\ log period for the
best samples available for each cluster (see
\S\ref{sec:bestsample}).  There is no correlation here, but keep
in mind that stars with periods are X-ray bright (see previous
figure).  The dashed line is the relationship between 
\lx/\lbol\ found by Preibisch \etal\ (2005) for the ONC, and the
solid line is the approximate relationship found for main
sequence stars.  We do not see a similar correlation (either the
ONC or MS one) for either the Orion FF or NGC 2264. }
\label{fig:period}
\end{figure*}

\begin{figure*}[tbp]
\epsscale{0.5}
\plotone{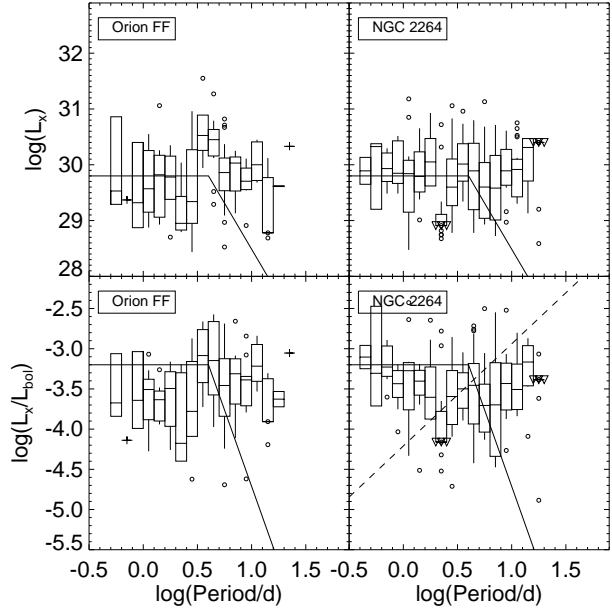}
\caption{The data from Fig.~\ref{fig:period} but in box plot
form.  We do not see a correlation similar to that from the ONC
(dashed line) or from the main sequence (solid line) for either
cluster.}
\label{fig:periodprime}
\end{figure*}

Figure~\ref{fig:period} plots log \lx\ and log \lx/\lbol\
against log period for our sample, and
Figure~\ref{fig:periodprime} shows the same data but in box plot
form.  There is no significant  trend in these plots, nor in
similar plots of X-ray emission against linear $P$, \vsini,  or
$1/P$.  Upper limits are found at all periods and rotation
rates.  The current data do not support as strong a correlation
between period and  fractional X-ray luminosity as the one
reported for the ONC by  Preibisch \etal\ 2005b.  We also do not
find the same correlations as can be found in main sequence
stars.


\section{Disk and Accretion Effects}
\label{sec:disks}


\begin{figure*}[tbp]
\epsscale{0.5}
\plotone{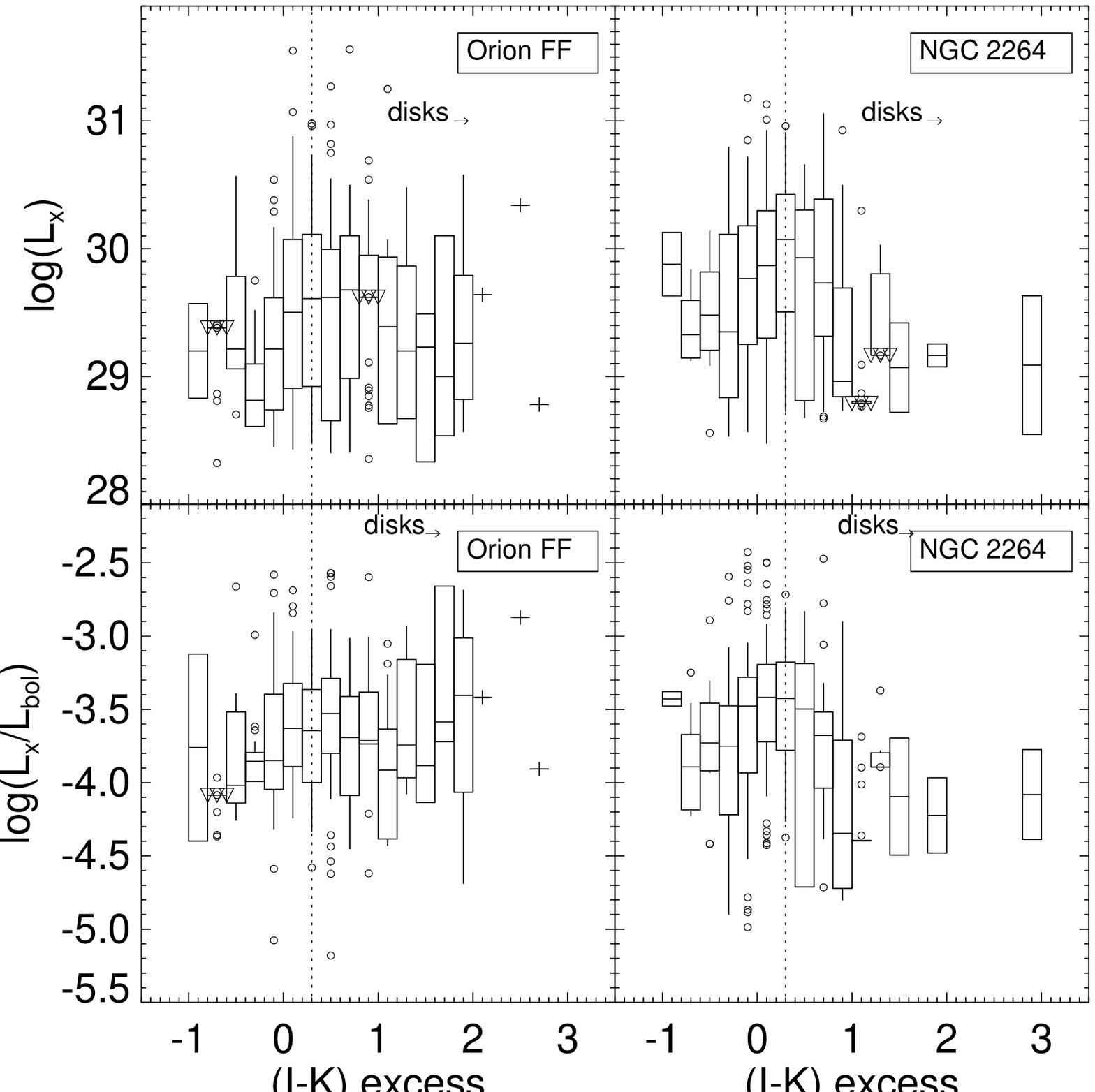}
\caption{Log \lx\ and log \lx/\lbol\ vs.\ \ik\ excess disk candidates for the
best samples available for each cluster (see \S\ref{sec:bestsample}).
Disk candidates are redder than 0.3 mags (to the right of the dotted
line; see text). }
\label{fig:lxikx}
\end{figure*}

\begin{figure*}[tbp]
\epsscale{0.5}
\plotone{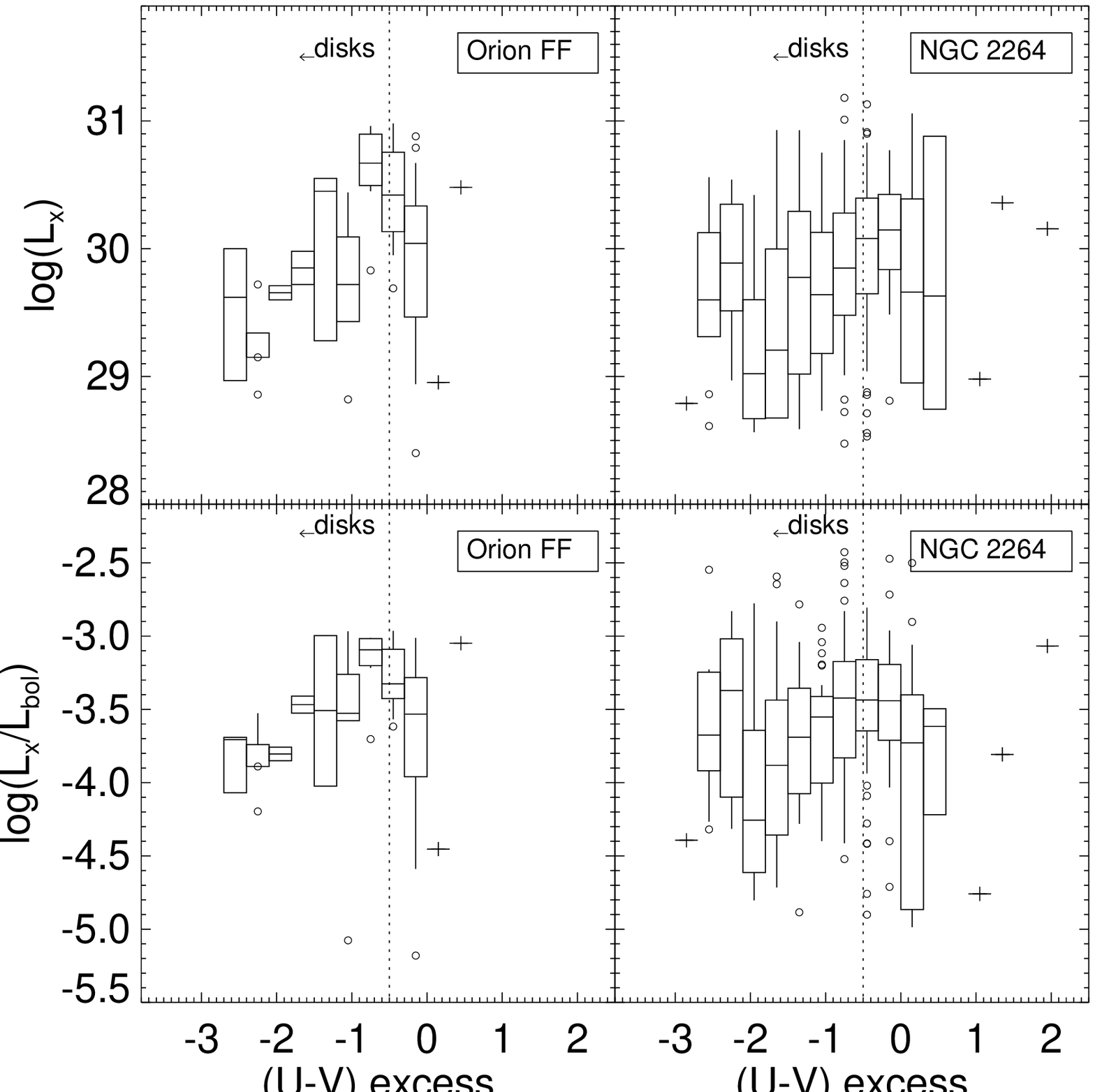}
\caption{Log \lx\ and log \lx/\lbol\ vs.\ \uv\ excess disk candidates for the
best samples available for each cluster (see \S\ref{sec:bestsample}).
Disk candidates are bluer than $-$0.5 mags (to the left of the dotted
line; see text). }
\label{fig:lxuvx}
\end{figure*}


Two processes associated with disk accretion can, in principle, 
affect the observed X-ray properties of solar-type PMS stars:
(1) local heating of the stellar photosphere at the
`footprints' of magnetospheric funnel  flows, where highly
supersonic ($v >$ 200 km sec$^{\-1}$) gas lands on the  stellar
surface (\eg\ Kastner \etal\ 2002; Schmitt \etal\ 2005), 
producing shocks, local heating and, as a result soft X-rays;
and (2) absorption of  keV coronal X-ray emission  by gas
associated with accretion-driven outflows (Walter \& Kuhi,
1981). The first of these  processes would qualitatively produce
an {\em increase} in X-ray emission above coronal activity
levels as disk  accretion rates increase. Quantitative estimates
of the magnitude of the expected increase are not  available,
owing primarily to the complexity of the detailed radiative
transfer studies needed to estimate the X-ray flux emerging
from accretion spots. The second process would produce a {\em
decrease} of X-ray emission with increasing accretion rate,
since the column density, and hence the X-ray optical  depth of
outflowing material, is expected  to increase with increasing
accretion rate. For isotropic outflows, significant 5 keV
absorption is expected to occur for mass outflow rates of
$10^{-8}$ \msun\ yr$^{-1}$ or mass accretion rates of
$\sim10^{-7}$ \msun\ yr$^{-1}$.

There has been a great deal of work attempting to uncover
relationships between circumstellar disks, and/or disk
accretion, and/or X-ray emission (either log \lx\ or log
\lx/\lbol).   Table~\ref{tab:diskliterature} collects the
conclusions from large surveys of X-ray emission for accreting
and non-accreting PMS stars. Kastner \etal\ (2002); and Schmitt
\etal\ 2005 find evidence of enhanced X-ray emission associated
with transient increases in disk accretion rate, thus 
suggesting that at least some X-ray emission may be associated
with accretion as opposed to coronal activity. In ensemble,
there is no evidence of {\em excess} X-ray emission for
accreting PMS stars. Rather, the trend, if any, is for an
anti-correlation between the presence of an accretion disk and
total X-ray luminosity.

Walter \& Kuhi (1981) were the first to report an inverse
correlation between soft X-ray flux and H$\alpha$  emission for
classical T Tauri stars. Working under the assumption that the
observed  equivalent width of H$\alpha$ serves as a surrogate
for mass outflow rate, they argued that the apparent decrease of
X-ray emission with increasing H$\alpha$ equivalent width
reflected the  effects of X-ray absorption by outflowing gas. 
More recently, Flaccomio \etal\ (2003b) find a similar result: a
larger fraction of stars with active accretion signatures, there
measured by \ion{Ca}{2} emission, have fainter \lx\ values than
non-accretors when stars of similar masses are compared. 
Stassun \etal\ (2004) also report a similar result for a sample
of young solar-like PMS stars, particularly
for those stars with $M<$ 0.5 \msun.  Preibisch \etal\ (2005b) find
significantly lower X-ray emission for accreting as opposed to
non-accreting PMS stars (as selected on  the basis of
\ion{Ca}{2} triplet emission equivalent width), for the
restricted mass range 0.3-0.5 \msun.  They also find that the
non-accretors show a better-defined correlation between \lx\ and
\lbol\ than the accretors, that the median value of  \lx/\lbol\
is nearly a factor of three lower for accretors, and that there
is a weak anti-correlation of the fractional X-ray luminosity
with accretion rate.  In looking for correlations of X-ray
emission with the presence/absence of accretion disks, but for
which there is no direct evidence of accretion, Preibisch \etal\
(2005b) find, somewhat surprisingly, that there are significant
differences in \lx/\lbol\ for stars with and without near-IR
\ik\ excesses, but no differences for stars with and without
$K-L$ excesses; $K-L$ is usually considered to be a more
reliable disk indicator.  

In our particular case, we have three different disk and/or
accretion indicators (\ik, \hk, and \uv\ excesses) for both
clusters, and a fourth indicator (H$\alpha$ equivalent widths)
for NGC 2264.  We note that the H$\alpha$ measurements were only
obtained for stars with classification spectra in NGC 2264. 
Most of these spectra were obtained as part of the studies by
Rebull \etal\ (2002) and Makidon \etal\ (2004), where the goal
was to classify stars with known periods.  Therefore, the sample
of stars with H$\alpha$ is likely to be biased toward stars that
are brighter in X-rays (see Figure~\ref{fig:pbiaslxlbol}).

The numbers of stars available from the best possible sample are
summarized in Table~\ref{tab:numdisks}.  While \ik\ and \hk\
excesses indicate the presence of either passive or actively
accreting circumstellar disks, \uv\ excesses and H$\alpha$ arise
from active accretion; \uv\ excesses have already been
translated directly to mass accretion rates for many of our
targets (Rebull \etal\ 2000, Rebull \etal\ 2002).  

Based on previously published results, we might expect that any
correlations between total and fractional X-ray luminosity and
stellar properties are likely to be weak (and in fact they
are).  We show plots for two of our available diagnostics of
disks and/or accretion in Figures~\ref{fig:lxikx} and
\ref{fig:lxuvx}.  Note that, for the near-IR disk indicators,
redder colors suggest disks, whereas for \uv, bluer colors
suggest accretion. The placement of the disk cutoffs is
discussed in Rebull \etal\ (2002, 2000), along with comparison
of the disk indicators for the same stars.

There is a very large amount of scatter in these two figures,
plus their companions (\hk\ and H$\alpha$) which are not shown,
although the scatter is somewhat smaller in the plots with
log \lx/\lbol\ than in the plots against log \lx.  For this
reason, we have chosen to show only the box plot representations
of these data.  

No statistically significant or consistent trends are seen in
Fig.~\ref{fig:lxikx} or \ref{fig:lxuvx} (or in similar plots
with  the other available disk indicators). In stars {\em
without} evidence for disks, we find examples of stars in which
\lx\ is high and log \lx/\lbol\ is close to $-$3, but comparable
large values are not seen in stars {\em with} disks or active
accretion.  The upper bound on both the total and fractional
X-ray luminosity appears to decrease with increasing  \ik\
excess for NGC 2264 (but not the FF) and in increasing \uv\
excess for the FF (but not NGC 2264).  The X-ray fluxes do not
appear to depend on \hk\ excess for either cluster or in
H$\alpha$ for NGC 2264.  These confusing, inconsistent results
suggest that if there is any trend, it is subtle, and, despite
the numbers of points we have (see Table~\ref{tab:numdisks}), we
require many more data points to securely detect it.

\begin{deluxetable}{lllll}
\tablecaption{Summary of numbers of stars available with and
without disks
\label{tab:numdisks}}
\tablewidth{0pt}
\tablehead{
\colhead{} & \multicolumn{2}{c}{Orion FF} &
\multicolumn{2}{c}{NGC 2264} \\
\colhead{disk indicator} & \colhead{non-disk} & 
\colhead{disk} & \colhead{non-disk} & 
\colhead{disk} \\
\colhead{} & \colhead{ detections (limits)} 
& \colhead{ detections (limits)}& \colhead{ detections (limits)}
& \colhead{ detections (limits)}}
\startdata
\ik\ excess & 129 (38) & 109 (37) & 226 (42) & 63 (21) \\
\hk\ excess & 171 (56) & 64 (15) & 236 (56) & 52 (15) \\
\uv\ excess & 28 (4) & 25 (3) & 87 (14) & 151 (48)\\
H$\alpha$ eqw & \nodata & \nodata & 77 (8) & 46 (12) \\
\enddata
\end{deluxetable}


\begin{figure*}[tbp]
\epsscale{0.5}
\plotone{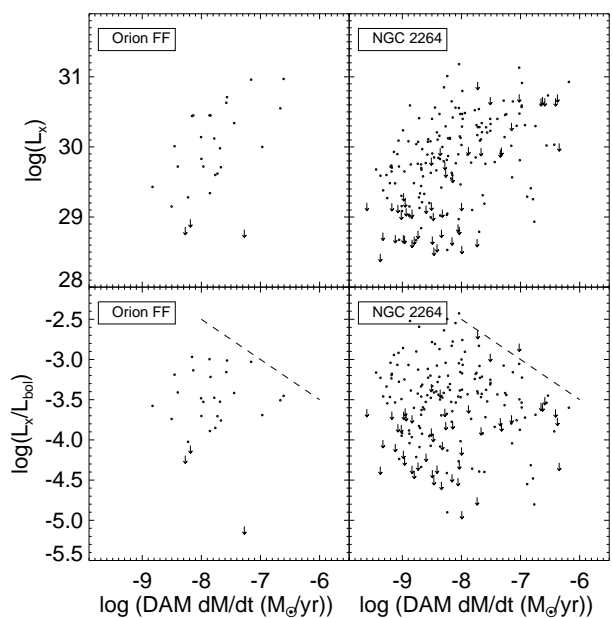}
\caption{Log \lx\ and log \lx/\lbol\ vs.\ \mdot\ (derived using
DAM masses) for the best samples available for each cluster (see
\S\ref{sec:bestsample}).  There is a trend in \lx, which is a
rediscovery of the mass-dependent effects of \lx\ found in
Fig~\ref{fig:mass} and of \mdot\ found in Rebull \etal\ (2000). 
There is no trend in \lx/\lbol, but the dashed lines trace a
trend with the highest \mdot\ values.}
\label{fig:mdot}
\end{figure*}


Figure~\ref{fig:mdot} looks for correlations explicitly with log
\mdot\ as derived from the \uv\ excess as in Rebull \etal\
(2000, 2002).  The data in the Orion FF are too sparse to reach
any conclusion. For NGC 2264, however, there may be evidence of
a trend: the upper envelope of \lx/\lbol\ is near $10^{-3}$ for
stars with accretion rates $<10^{-8}$ \mdot\ yr$^{-1}$, but
progressively decreases by nearly a factor of 10 for stars with
higher accretion rates --  a result in concordance with
previously reported anti-correlations between X-ray luminosity
and accretion for solar-like PMS stars.  

It is perhaps surprising that there are no clear correlations of
X-ray emission  with either the presence/absence of accretion
disks, or the magnitude of accretion rate apparent  from our
data, particularly given the plausibility of proposed X-ray
production and absorption  processes. Perhaps the intrinsic
range in X-ray emission properties is so large that trends are
difficult to  discern, even with samples as large as several
hundred stars. Alternatively, it may be that  multiple
mechanisms are at work simultaneously, and sorting among them
may require a somewhat  finer characterization of X-ray
properties than  Lx -- the single property thus far used in 
searching for accretion-related correlations.

We anticipate that the COUP data, in combination with the Orion
Treasury program (Robberto \etal\ 2005) and the Spitzer data in
Orion (Megeath \etal\ 2006) will provide considerably more
insight into the physical processes  accompanying disk
accretion. In particular, the Treasury Program will provide
robust estimates of  ultraviolet excess emission for a large
sample, which in combination with extant optical spectroscopic
studies yielding $R$ and \teff\ estimates, should provide
accretion rate estimates for a large (perhaps 1000 stars) sample
of accreting PMS stars. The deep COUP exposures of the ONC (20
$\times$ the length of the  Chandra observations reported here) 
should provide well-characterized X-ray spectral  energy
distributions from 0.5 to 5 keV -- {\em sine qua non} for
sorting the relative contributions of soft X-ray  emission that
may arise from accretion, harder coronal X-ray emission and the
effects of absorption arising in mass outflows.

\section{Conclusions}
\label{sec:concl}

We have studied the X-ray properties of stars in the Orion FF
and NGC 2264. With ages of  $\sim$1 Myr and $\sim$3 Myr
respectively, both regions contain stars that are older than
most stars in the ONC. Rotation periods are available for a
relatively large fraction of the PMS population in these
regions. Furthermore, NGC 2264 contains a significant number of
stars that have made the transition to radiative tracks. These
samples thus enable study of evolution- and rotation-driven
changes in X-ray properties over the age range 1-5 Myr.

The data for the Orion FF and NGC 2264, when combined with the
much more extensive data from the COUP survey of the ONC,
establish a number of clear patterns in the behavior of X-ray
emission in PMS stars. In all three regions, log \lx\ for stars
on convective tracks correlates  positively with both log \lbol\
and also with stellar mass, not surprisingly since log \lbol\ is
strongly correlated with mass.   As Preibisch  \etal\ (2005b)
point out, a strong correlation between X-ray emission and
stellar properties argues that the X-ray emission probably
originates  primarily in compact (coronal?) structures
associated with the star.  We confirm as well that the level of
X-ray emission is directly related to stellar structure in that
the level of emission drops on average by $\sim$10 when PMS
stars with masses in the range $\sim$ 1-2 \msun\ make the
transition from fully convective to radiative tracks, where
stars of about solar mass have radiative cores and convective
envelopes.

The fractional X-ray luminosity, log \lx/\lbol, is not strongly
correlated with mass for stars on convective tracks; the scatter
in the relationships of both log \lx\ and log \lx/\lbol\ with
mass is large. The physical reasons for this large scatter
remain unclear.  Age is apparently not a large contributor;
while there is evidence the ONC  has nearly twice as many stars
with extremely high X-ray emission (log \lx/\lbol\ $\geq-2.9$) 
as the Orion FF and NGC 2264, we see little change in {\em
median} values of either log \lx\ or log \lx/\lbol\ during the
first $\sim$5 Myr of evolution down convective tracks.   In
sharp contrast to  main sequence stars, we find no correlation
between log \lx\ and log \lx/\lbol\ with $P$ or \vsini\ for 
stars on convective tracks, although more \vsini\ data are
needed in NGC 2264.  Preibisch \etal\ (2005) have argued that
X-ray variability at typical levels is likely to be too small to
account for the scatter in \lx/\lbol.

In both the Orion FF and NGC 2264 we find, in agreement with 
earlier studies of the ONC, that stars with larger values of log
\lx/\lbol\ are  more likely to exhibit spot-modulated periods. 
Approximately $3/4$ of the stars with log \lx/\lbol$>-3$ are
periodic, while while only about $1/3$ of the stars with log
\lx/\lbol$<-4$ have periods, and this difference does not appear
to be the result of any biases in the samples.  It may well be
that, as Stassun \etal\ (2004)  have suggested, more active
stars have larger and/or more organized magnetic spot coverage
and that periodic variations are more easily measured in such
circumstances.

We have searched for a correlation between log \lx\ and log
\lx/\lbol\  and  1) near-IR excess, which is a circumstellar
disk indicator; 2) UV excess and H$\alpha$, both of which are
indicators of accretion; and 3) the mass accretion rate, as
inferred from the UV excess.  There is substantial scatter in 
all of these relationships, and no clear trends emerge. 
Discovery of the relationship, if any, between accretion and
X-ray emission awaits the combination of COUP results, which
provides measures of X-ray luminosity and X-ray spectral energy
distributions, with the corresponding results from the ongoing
HST Orion Treasury program, which promises to provide accurate
estimates of accretion rates for large samples of PMS stars, and
the Spitzer data in Orion, which will provide more accurate
mid-IR disk indicators.

\acknowledgements 
We would like to thank J.\ Najita for extensive discussions
regarding the power of X-ray observations to test competing wind
models.   E.\ F., G.\ M. and S.\ S.\ acknowledge support from
INAF and MIUR-PRIN grants. This research has made use of data
products from the Two Micron All-Sky Survey (2MASS), which is a
joint project of the University of Massachusetts and the
Infrared Processing and Analysis Center, funded by the National
Aeronautics and Space Administration and the National Science
Foundation.  These data were served by the NASA/IPAC Infrared
Science Archive, which is operated by the Jet Propulsion
Laboratory, California Institute of Technology, under contract
with the National Aeronautics and Space Administration.  The
research described in this paper was partially carried out at
the Jet Propulsion Laboratory, California Institute of
Technology, under contract with the National Aeronautics and
Space Administration.


\begin{thebibliography}{}

\bibitem[Alcala(2000)]{alcala00} Alcala, J., \etal\ 2000, A\&A, 353, 186
\bibitem[Bouvier(1990)]{bouvier90} Bouvier, J. 1990, AJ, 99, 946
\bibitem[Casanova(1995)]{casanova95}  Casanova, S., \etal\ 1995, ApJ, 439, 752
\bibitem[Chandra1950]{chandra1950}Chandrasekhar, S. \& Muench, G. 
1950, ApJ, 111, 142 
\bibitem[Dahm \& Simon(2005)]{dahm05} Dahm, S., \& Simon, T. 2005, AJ, 129,
829
\bibitem[Damiani \& Micela(1995)]{damiani95} Damiani, F., \& Micela, G.
1995, ApJ, 446, 341
\bibitem[D'Antona \& Mazzitelli(1994)]{dam94}D'Antona, F., \& Mazzitelli,
I.\ 1994, \apjs, 90, 467 (DAM)
\bibitem[D'Antona \& Mazzitelli(1998)]{dam98}D'Antona, F., \& Mazzitelli,
I.\ 1998, http://www.mporzio.astro.it/$\sim$dantona/prems.html (DAM)
\bibitem[eisner05]{eisner05} Eisner, J., \etal\ 2005, \apj, 623, 952
\bibitem[favata]{favata}Favata, F., \etal\ 2005, \apjs, 160, 469
\bibitem[Feigelson(1993)]{feigelson93}  Feigelson, E., \etal, 1993, ApJ, 416, 623
\bibitem[Feigelson(1999)]{feigelson99}  Feigelson, E. \& Montmerle, T.
1999, ARAA, 37, 363
\bibitem[Feigelson(2002)]{feigelson02}  Feigelson, E., \etal, 2002, ApJ, 574, 258
\bibitem[Feigelson(2003)]{feigelson03}  Feigelson, E., \etal, 2003, ApJ, 584, 911
\bibitem[Feigelson(2004)]{feigelson04}  Feigelson, E., \etal, 2004, in
Proceedings of ``Cool Stars 13,'' July 2004, Hamburg, Germany
\bibitem[FeigelsonLawson(2004)]{feigelsonlawson04}  Feigelson, E., \& Lawson, 
2004, ApJ, 614, 267
\bibitem[Flaccomio(1999)]{flaccomio99}Flaccomio, E., Micela, G.,
Sciortino, S, Favata, F., Corbally, C., \& Tomaney, A. 1999, A\&A, 345, 521
\bibitem[Flaccomio(2000)]{flaccomio00}  Flaccomio \etal\ 2000, A\&A, 355, 651
\bibitem[Flaccomio(2003a)]{flaccomio03a}  Flaccomio \etal\ 2003a, A\&A, 397, 611 
\bibitem[Flaccomio(2003b)]{flaccomio03b}  Flaccomio \etal\ 2003b, ApJ, 582, 398 
\bibitem[Flaccomio(2003c)]{flaccomio03c}  Flaccomio \etal\ 2003c, A\&A, 402, 277 
\bibitem[Flaccomio(2005)]{flaccomio05}  Flaccomio \etal\ 2005, Mem. S.
A. It., 76, 279
\bibitem[Flaccomio(2005b)]{flaccomio05b}  Flaccomio \etal\ 2006, in
preparation 
\bibitem[Gagne(1994)]{gagne94}  Gagne, M., \& Caillault, J.-P., 1994, ApJ, 437, 361
\bibitem[Gagne(1995)]{gagne95}  Gagne, M., Caillault, J.-P., \& Stauffer, J., 1995, ApJ, 445, 280
\bibitem[Gaige1993]{gaige93} Gaige, Y. 1993, A\&A, 269, 267
\bibitem[Getman(2002)]{getman02}  Getman, K., \etal, 2002, ApJ, 575, 354
\bibitem[Getman(2005)]{getman05}  Getman, K., \etal, 2005, ApJS, 160,
352
\bibitem[Grosso(2000)]{grosso00}  Grosso \etal\ 2000, A\&A, 359, 113
\bibitem[Guedel]{guedel} Guedel, M., \etal\ 2005, \apjl, 626, 53
\bibitem[Gullbring(1998)]{gullbring98} Gullbring, E., \etal\ 1998,
\apj, 492, 323
\bibitem[Hartmann]{hartmann}Hartmann, L. 2001, ApJ 121, 1030
\bibitem[Hillenbrand(97)]{hillenbrand97} Hillenbrand, L., 1997,
AJ, 113, 1733
\bibitem[Hillenbrand(2004)]{hillenbrand04} Hillenbrand, L., \& White, R.,
2004, ApJ, 604, 741
\bibitem[Imanishi(2001)]{imanishi01}  Imanishi, K., \etal, 2001, ApJ, 557, 747
\bibitem[Kastner(1997)]{kastner97} Kastner, J., \etal\ 1997, Science, 277, 67
\bibitem[Kastner(2002)]{kastner02} Kastner, J., \etal\ 2002, ApJ, 567, 434
\bibitem[Kastner(2005)]{kastner05} Kastner, J., \etal\ 2005, ApJS,
160, 511
\bibitem[KenyonHartmann(1995)]{KH95}Kenyon, S., \& Hartmann, L.,
1995, ApJS, 101, 117
\bibitem[K\"onigl (2000)]{konigl00} K\"onigl, A. \& Pudritz, R., 2000,
in Protostars and Planets IV, (Tucson: University of Arizona
Press; eds Mannings, V., Boss, A.P., Russell, S. S.), p. 759
\bibitem[K\"onigl (1991)]{konigl91} K\"onigl, A. 1991, \apj, 370, L39
\bibitem[K\"onigl (1989)]{konigl89} K\"onigl, A. 1989, \apj, 342, 208
\bibitem[Lamm(2004)]{lamm2004}Lamm, M., \etal\ 2004, A\&A, 417, 557
\bibitem[Lawson(1996)]{lawson96}  Lawson \etal\ 1996, MNRAS, 280, 1071
\bibitem[Makidon(2004)]{makidon04} Makidon, R., \etal\ 2004, AJ, 127, 2228
\bibitem[Megeath]{megeath}Megeath, T., 2006, in prep
\bibitem[Messina(2003)]{messina03} Messina, S., Pizzolato, N., Guinan, E.,
Rodono, M. 2003, \aap, 410, 671
\bibitem[Micela(1996)]{micela96} Micela, G., et al. 1996, ApJS, 102,
75
\bibitem[MorrisonMcCammon]{morrison}Morrison, R., \& McCammon,
D., 1983, ApJ, 270, 119
\bibitem[Muzerolle2003]{muzerolle03} Muzerolle, J, \etal\ 2003, \apj,
597, 149
\bibitem[Najita2003]{najita03}Najita, J., \etal\ 2003, \apj, 589, 931
\bibitem[Neuhauser(1995)]{neuhauser95}  Neuhauser, R., \etal\ 1995, A\&A, 297, 391 
\bibitem[Ozawa(2005)]{ozawa05}  Ozawa, H., Grosso, N., Montmerle, T. 2005,
A\&A, 429, 963
\bibitem[Park(2000)]{park2000}Park, B.-G., et al. 2000, AJ, 120, 894
\bibitem[Pizzolato(2003)]{pizzolato03} Pizzolato, N., Maggio, A., Micela,
G., Sciortino, S., Ventura, P., 2003, A\&A, 397, 147
\bibitem[Preibisch(1997)]{preibisch97} Preibisch, T., 1997, A\&A, 320, 525
\bibitem[Preibisch(2002)]{preibisch02}  Preibisch, T. \& Zinnecker, H. 2002, AJ, 123, 1613
\bibitem[Preibisch(2005)]{preibisch05}  Preibisch, T.  2005, A\&A, 428, 569
\bibitem[Preibisch(2005b)]{preibisch05b}  Preibisch, T.  \etal\ 2005,
ApJS, 160, 401
\bibitem[Preibisch(2005c)]{preibisch05c} Preibisch, T., \& Feigelson,
E., 2005, \apjs, 160, 390
\bibitem[press]{press}Press, W.\ H., Teukolsky, S.\ A., Vetterling, W.\
T., Flannery, B.\ P., 1992, {\it Numerical Recipes in C}, 2nd edition
(Cambridge: Cambridge University Press)
\bibitem[Ramirezetal(2004a)]{ramirez04a} Ramirez, S., Rebull, L., Stauffer,
J., Hearty, T., Hillenbrand, L., Jones, B., Makidon, R., Pravdo, S., Strom,
S., Werner, M., 2004a, AJ, 127, 2659 
\bibitem[Ramirezetal(2004b)]{ramirez04b} Ramirez, S., Rebull, L., Stauffer,
J., Strom, S., Hillenbrand, L., Hearty, T., Kopan, E., Pravdo, 
S., Makidon, R., Jones, B., 2004b, AJ, 128, 787 
\bibitem[Rebull(2001)]{rebull01}  Rebull, L. 2001, \aj, 121, 1676 (R01)
\bibitem[Rebull et~al.(2000)]{rebull00}  Rebull, L., Hillenbrand, L. A., 
Strom, S. E., Duncan, D. K., Patten, B. M., Pavlovsky, C. M., Makidon,
R. B., \& Adams, M. T. 2000, \aj, 119, 3026
\bibitem[Rebull et~al.\ (2002)]{rebull02} Rebull, L., Makidon, R. B., 
Strom, S. E., Hillenbrand, L. A., Birmingham, A., Yagi, H., Jones, B. F., 
Adams, M. T., \& Patten, B. M. 2002, \aj, 123, 1528 (RMSH02)
\bibitem[Rebull et~al.\ (2002b)]{rebull02b} Rebull, L., Wolff, S., Strom,
S, \& Makidon, R. B. 2002b, AJ, 124, 546
\bibitem[Rebull et~al.\ (2004)]{rebull04} Rebull, L., Wolff, S.,
\& Strom, S. E., 2004, AJ, 127, 1029
\bibitem[Rhodeetal(2001)]{rhode2001} Rhode, K., et al., 2001, \aj,
122, 3258
\bibitem[Robberto2005]{robberto05} Robberto, M., 2005, in  ``Star
Formation in the Era of Three Great Observatories,'' held July 13-15,
2005, in press.
\bibitem[Schmitt04]{schmitt04}Schmitt, J., and Liefke, C., 2004, \aa,
417, 651
\bibitem[Schmitt05]{schmitt05}Schmitt, J. \etal\ 2005, A\&A, 432, L35
\bibitem[Shu(1987)]{shu1987} Shu, F. \etal\ 1987, ARAA, 25, 23
\bibitem[Shu(2000)]{shu2000}  Shu, F., Najita, J., Shang, H., \& Li, Z.-Y. 
2000, in Protostars and Planets IV, ed. V. Mannings, A. P. Boss \& 
S. S. Russell (Tucson: University of Arizona Press), p. 789
\bibitem[SiciliaAguilar(2005)]{sicilia}Sicilia-Aguilar, A., \etal, 2005,
\aj, 129, 363
\bibitem[Siess et al.(2000)]{sdf}Siess, L., Dufour, E., \& Forestini, M.\
2000, \aap, 358, 593, and
$<$http://www-laog.obs.ujf-grenoble.fr/activites/starevol/evol.html$>$
(SDF)
\bibitem[Sung(1997)]{sung97}Sung, H., Bessell, M., \& Lee, S.-W. 1997, AJ,
114, 2644
\bibitem[Sung(2004)]{sung04}Sung, H., Bessell, M., \& Chun, M.-Y. 2004, AJ,
128, 1684
\bibitem[Stassun(2004)]{stassun04}Stassun, K. \etal\ 2004, AJ, 127, 3537
\bibitem[Stauffer(1994)]{stauffer94}Stauffer, J. R., \etal\ 1994, ApJS, 91,
625
\bibitem[strom94]{strom94}Strom, S., 1994, ASPC, 64, 211
\bibitem[Stelzer(2000)]{stelzer00}  Stelzer, B. \etal\ 2000, A\&A, 356, 949
\bibitem[Stelzer(2001)]{stelzer01}  Stelzer, B. \& Neuhauser, R., 2001, A\&A, 377, 538
\bibitem[Tsujumoto(2002)]{tsujimoto02}  Tsujimoto, M., \etal\ 2002, ApJ, 566, 974
\bibitem[Walter(1981)]{walter81} Walter, F., \& Kuhi, L., 1981, ApJ,
250, 254
\bibitem[Wichmann(2000)]{wichmann00} Wichmann, R., \etal\ 2000, A\&A, 359, 181
\end{thebibliography}
\end{document}